\documentclass[preprint, 3p, sort&compress]{elsarticle}

\usepackage{hyperref}
\usepackage{amsmath}
\usepackage{amsfonts}
\usepackage{tikz}
\usepackage{epigraph}
\usepackage{subcaption}
\usepackage{eurosym}

\newcommand{\Exp}{\mathbb{E}}

\usetikzlibrary{arrows.meta}
\tikzset{%
	>={Latex[width=2mm,length=2mm]},
	base/.style = {rectangle, rounded corners, draw=black,
		minimum width=2cm, minimum height=.75cm,
		text centered,font=\rmfamily\footnotesize},
	output/.style = {base, fill=blue!15},
	input/.style = {base, fill=green!15},
	process/.style = {base, fill=black!15},
	truth/.style = {base, fill=white}
}

\journal{arXiv}

\begin{document}

\begin{frontmatter}

\title{Sensitivity measures for engineering and environmental decision support}



\author[ERA,MDSI]{Daniel Straub\corref{mycorrespondingauthor}}
\cortext[mycorrespondingauthor]{Corresponding author}

\author[Eracons]{Wolfgang Betz}
\author[ERA]{Mara Ruf}
\author[ERA]{Amelie Hoffmann}
\author[NAGRA]{Angela Landgraf}
\author[ERA]{Lea Friedli}
\author[ERA]{Iason Papaioannou}

%
\address[ERA]{Technical University of Munich, Engineering Risk Analysis Group}
\address[MDSI]{Munich Data Science Institute}
\address[Eracons]{Eracons}
\address[NAGRA]{NAGRA}

\begin{abstract}
Information value, a measure for decision sensitivity, can provide essential information in engineering and environmental assessments. It quantifies the potential for improved decision-making when reducing uncertainty in specific inputs. By contrast to other sensitivity measures, it admits not only a relative ranking of input factors but also an absolute interpretation through statements like ''Eliminating the uncertainty in factor $A$ has an expected value of $5000$ Euro''. In this paper, we present a comprehensive overview of the information value by presenting the theory and methods in view of their application to engineering and environmental assessments. We show how one should differentiate between aleatory and epistemic uncertainty in the analysis. Furthermore, we introduce the evaluation of the information value in applications where the decision is described by a continuous parameter. The paper concludes with two real-life applications of the information value to highlight its power in supporting decision-making in engineering and environmental applications.


\end{abstract}

\begin{keyword}
Sensitivity analysis \sep uncertainty \sep value of information \sep decision support \sep flood risk 
\end{keyword}

\end{frontmatter}

\graphicspath{{Figures/}}



\section{Introduction}
\label{sec:Intro}

Models are the basis for decision support in engineering and environmental applications. However, 
models and their parameters have uncertainties. It is recognized that good decision support should acknowledge and adequately communicate these uncertainties when they are relevant \cite{ascough_future_2008,laniak_integrated_2013,beven2018}. One important element of this is a global sensitivity analysis, which tries to quantify the effect of uncertainties on the quantities of interest predicted with the model, i.e., the model outcome.

Formally, let $Y=\mathcal{Y}(\mathbf{X})$ be the outcome of a model $\mathcal{Y}$, with uncertain input parameters $\mathbf{X}$, which are modeled as random variables. In alignment with the sensitivity analysis literature, we refer to the elements of $\mathbf{X}$ as factors. Global sensitivity analysis (GSA) methods provide insights into how the properties of $\mathbf{X}$ affect the outcome $Y$. 
GSA methods include (but are not limited to) variance-based \cite{sobol1993sensitivity,Jansen1999,Saltelli2010} and distribution-based \cite{Chun2000,Borgonovo2007,Borgonovo2016a} sensitivity measures. 

Sensitivity measures are (seemingly) easy to interpret in a relative sense, such as in ''factor $A$ is more important than factor $B$''. This has led to the widespread use of sensitivity analysis in science and engineering, in most cases without much reflection on the meaning and definition of different sensitivity measures \cite{Saltelli2019}. In particular, the following two challenges are not properly addressed in many engineering and environmental sensitivity analyses. 

The first challenge is the choice of the measure. Many different sensitivity measures have been proposed in the literature \cite{saltelli2008global,Iooss2015a,Borgonovo2016,pianosi_sensitivity_2016}. It is good practice to distinguish between sensitivity measures for factor prioritization, for factor fixing and for factor mapping. However, within these categories, there are different sensitivity measures one can select. Each measure is associated with a different definition of what an ''important factor'' is.
In practice, the choice of a specific measure is often guided by what measure the analyst is aware of or what is traditionally used in a specific application domain, rather than informed by a deeper understanding of the meaning attached to these measures. 

The second challenge is that engineering and environmental decision-making calls for a statement about sensitivity in an absolute sense and not just a relative ranking of factors. 
When presenting the results of an assessment, what truly matters to the decision-maker is how uncertainties in the assessment affect its conclusions. In the words of Felli and Hazen \cite{Felli1998a}, the ''decision sensitivity''.

To illustrate the decision sensitivity, consider the following motivating example: In Switzerland, three possible underground storage sites were identified for nuclear waste repositories in past studies \cite[Section~1.3]{nagra2021}. Detailed assessments of these three sites were performed to decide which one is best suited to host the repository, whereby a number of criteria were considered. 
One of these criteria is Erosion, which was assessed through the probability that the repository is exposed within the next one million years. Such an adverse event may compromise the function of the natural barrier, i.e., the clay rock hosting the deep geological repository. 
Detailed geological assessments were carried out to evaluate the future behavior of the erosion system.
However, even with an extensive state-of-the-art assessment, involving experts from different domains, uncertainty remains on the resulting probability of exposure. To utilize resources efficiently, one would like to know how these uncertainties affect the decision, namely the selection of the site among the three possible alternatives. Some factors may influence the probability substantially but not change the decision from one site to another, while other factors with a smaller effect on the probability might be more likely to lead to a change in the decision. Further analysis should then focus on these factors. Hence, the sensitivity of the model to different factors should be measured in a way that reflects this decision sensitivity. 

Decision sensitivity measures have existed for quite some time but have received little attention in engineering and environmental science practice. In the late 1990s, Felli and Hazen \cite{Felli1998a} proposed the use of the expected value of partial perfect information (EVPPI) as a decision sensitivity measure. Around the same time, P{\"o}rn \cite{Porn1997} proposed the same measure in the context of probabilistic safety assessment. The EVPPI, also called \textit{information value} \cite{Borgonovo2021}, quantifies how much is gained in the expected sense by full knowledge of an uncertain factor because of improved decision-making. 
Decision sensitivity has gained traction in the domain of medical decision-making, and most developments on the EVPPI have been made in that domain \cite[e.g.,][]{Felli1998a,Claxton1999,ades2004expected,baio2015probabilistic}. Fewer works have considered the EVPPI in the context of engineering or environmental decision-making, including \cite{Porn1997,yokota2004value,bates_value_2014,canessa_when_2015,goda2018decision,Fauriat2018,straub2022decision}. This is surprising to us, given that most models in the latter domains are directed toward decision support. 

This paper aims to foster the application of the information value as a sensitivity measure in engineering and environmental modeling. To this end, the paper provides an overview of the theory, models, and methods required for evaluating the information value in these applications. The main novelty of our contribution lies in this systematic overview. Additionally, we extend the existing methodology in some directions that are particularly relevant for engineering and environmental decision-making. In particular, we consider that different types of uncertainties, commonly referred to as aleatory and epistemic uncertainty \cite{der2009aleatory,hullermeier_aleatoric_2021}, are present in most engineering applications and we show the implications of this separation on the determination of the information value. We propose a novel method for normalizing information values that accounts for this separation. Finally, we extend the use of the information value to applications with continuous decision parameters, as are often encountered in engineering decision-making. 

The paper is intended to serve as a stand-alone paper that provides an overview of the theory and methods most relevant for utilizing information value in engineering and environmental applications. Section \ref{sec:Setup} introduces the uncertainty quantification (UQ) and decision analysis setup and notation required for decision sensitivity analysis. In \ref{sec:alea_epistemic}, we discuss the distinction between aleatory and epistemic uncertainty and how this can be considered in the UQ formulation. 
Section \ref{sec:WorkingExample} introduces a working example, which serves to illustrate the theory and methods throughout the paper. 
Section \ref{sec:decision_sensitivity} reviews the theory behind decision sensitivity and introduces a novel normalization that accounts for the separation between aleatory and epistemic uncertainty. 
In Section \ref{sec:sample_value} we present useful theory on the sample information value, which seems not to have been noticed outside the medical decision domain. 
In Section \ref{sec:decisions_utility}, we discuss the different classes of decisions in engineering and environmental applications and how to model them in the context of decision sensitivity analysis. The existing decision sensitivity literature focuses on discrete decisions; here, we extend the concept to continuous decisions.
In Section \ref{sec:Computing}, we discuss and propose effective and efficient computational methods to evaluate the information value for the different types of decision settings. 
Finally, in Sections \ref{sec:application_flood} and \ref{sec:application_nuclearWaste} we present two real-life applications of the information value to demonstrate the usefulness of the concept in engineering and environmental decision-making: The first one is a flood risk model of the Danube river that is used for cost-benefit analysis of flood mitigation measures. 
The second one is a geological model that is used 
to support the erosion assessment at three sites
for a nuclear waste repository in Switzerland.

\section{Model outcomes, decisions, utility and uncertainties}
\label{sec:Setup}

We previously introduced $Y=\mathcal{Y}(\mathbf{X})$ as the outcome of a model $\mathcal{Y}$, with uncertain input parameters $\mathbf{X}$, in agreement with standard UQ and sensitivity analysis. Often, $Y$ is called the quantity of interest (QoI).
Additionally, we define the decision parameter $a$, which specifies the decision to be taken. Often, $a$ is discrete; e.g., in the motivating example of Section \ref{sec:Intro}, $a$ would be one of three storage sites. In other cases, $a$ can be continuous, such as an engineering design parameter determining the capacity of a system. $a$ can be a vector, but in most practical cases it is sufficient to consider $a$ as a scalar parameter, and we will do so in the following. 

Importantly, the QoI can become a function of the decision $a$. 
For the 
example
of an erosion assessment of sites for a deep geological repository for nuclear waste, 
the QoI is the probability of exposure at the selected site, which obviously depends on the decision $a$ which site to choose. To account for this, we change the general notation to 
\begin{equation}\label{eq:general_model}
    Y(a)=\mathcal{Y}(\mathbf{X},a),
\end{equation}
to clarify that the QoI $Y$ is a function of the decision and that the model $\mathcal{Y}$ takes $a$ as an input.

This dependence on the decision has important implications for sensitivity analysis. In particular, the sensitivity of the QoI to specific inputs can be affected by the decision taken. Therefore, the sensitivity should not be considered independent of the decision. 

When considering decisions in the sensitivity analysis, it is necessary to define criteria for selecting decisions. 
In classical decision analysis \cite{Raiffa1961}, the utility function $u$ quantifies the preference of the decision maker for a specific combination of actions $a$ and outcomes $\mathbf{X}$.  
It is postulated that the optimal decision $a_{opt}$ is the one that maximizes the expected value of the utility\footnote{Alternatively, the statistics literature often considers the loss, defined as $L(\mathbf{X},a)=-u(\mathbf{X},a)$. Consequently, the goal is to minimize the expected loss \cite{robert_bayesian_2007}. This can be confusing to a wider audience, for which a loss would typically be associated with a physical or economic loss before considering any risk aversion. For risk-neutral decision-makers the two definitions of loss are, however, equivalent.}:
\begin{equation}
		\label{eq:prior_decision}
	a_{opt}=\arg \max_a \Exp_\mathbf{X}\left[u(\mathbf{X},a)\right].
\end{equation}
$\Exp_\mathbf{X}$ is the mathematical expectation with respect to $\mathbf{X}$. 

The model outcome $Y$ should be defined such that the utility function depends on $\mathbf{X}$ solely through $Y$. Therefore, the utility function can be written as
\begin{equation} \label{eq:define_utility_function_Y}
    u(\mathbf{X},a)=u'\left(\mathcal{Y}(\mathbf{X},a),a\right).
\end{equation}
One way to achieve this is to set the QoI to be the utility function, i.e., $\mathcal{Y}(\mathbf{X},a)=u(\mathbf{X},a)$, in which case $u'$ is just a projection onto its first argument. Alternative examples of functions $u'$ are provided in Sections \ref{sec:WorkingExample} and \ref{sec:decisions_utility}.

\subsection{Aleatory and epistemic uncertainty}\label{sec:alea_epistemic}
In many sensitivity analyses, it is relevant to distinguish between different types of uncertainty, even if this distinction is not often discussed in the sensitivity literature. Of particular relevance is the distinction between aleatory and epistemic uncertainty \cite{Der_Kiureghian2009aleatory,hullermeier_aleatoric_2021}.
\textit{Aleatory} uncertainty is also known as irreducible uncertainty or inherent randomness. 
Aleatory uncertainty is modeled by random variables $\mathbf{X}_a$ with associated probability distributions. An example of aleatory uncertainty is the maximum future water level at a levee, which is a random variable that follows an extreme value distribution. 
\textit{Epistemic} uncertainty, also known as model uncertainty or knowledge uncertainty, refers to quantities that are not inherently random, but are not known with certainty to the decision-maker at the time of making the decision. An example of epistemic uncertainty is the uncertainty in the capacity of the levee to withstand a certain water level. Another example is uncertainty in the parameters of the extreme value distribution describing the maximum water level, i.e., statistical uncertainty. Epistemic uncertainty is modeled by random variables $\mathbf{X}_e$.\footnote{Some authors argue that probabilistic models should only be used for aleatory uncertainties and propose alternative models for representing epistemic uncertainty \cite{ferson2004summary,beer2013imprecise}. We - as do most others \cite{o2004probability,der2009aleatory} - do not see a reason why epistemic uncertainty cannot also be represented by probabilistic models. This can be justified through a Bayesian viewpoint, which accepts probability as a subjective degree of belief, or simply as a pragmatic choice \cite{savage1956foundations,degroot1970optimal}.} It is $\mathbf{X}=\left[\mathbf{X}_a,\mathbf{X}_e\right]$.


Because there is nothing that the decision-makers can do to reduce aleatory uncertainty, quantifying the decision sensitivity with respect to these uncertainties is often irrelevant. 
Therefore, there are cases in which the sensitivity analysis should consider as inputs only the epistemic uncertainties $\mathbf{X}_e$  and as output the expected value of the utility function with respect to the aleatory uncertainties $\mathbf{X}_a$:
\begin{equation} \label{eq:utility_function_Epistemic}
    u(\mathbf{X}_e,a)=\Exp_{\mathbf{X}_a|\mathbf{X}_e}[u(\mathbf{X},a)],
\end{equation}
%
wherein $\Exp_{\mathbf{X}_a|\mathbf{X}_e}$ represents the expectation with respect to $\mathbf{X}_a$ conditional on $\mathbf{X}_e$.

If the QoI and the utility are not identical, then one might first evaluate the expected value of $\mathcal{Y}(\mathbf{X},a)$ with respect to $\mathbf{X}_a$ as
\begin{equation} \label{eq:general_model_epistemic}
    Y_e(a)=\mathcal{Y}_e(\mathbf{X}_e,a)=\Exp_{\mathbf{X}_a|\mathbf{X}_e} \left[\mathcal{Y}([\mathbf{X}_a,\mathbf{X}_e],a)\right],
\end{equation}
and then obtain the utility function of Eq. \ref{eq:utility_function_Epistemic} by inserting $\mathcal{Y}_e(\mathbf{X}_e,a)$ into Eq. \ref{eq:define_utility_function_Y}: $u(\mathbf{X},a)=u'\left(\mathcal{Y}_e(\mathbf{X}_e,a),a\right)$. 
Note that this is only admissible if the following holds:
\begin{equation}
	\label{eq:linear_utility_Xa}
    \Exp_{\mathbf{X}} \left[u'(\mathcal{Y}(\mathbf{X},a),a)\right] =  
    \Exp_{\mathbf{X}_e} \left[u'(\Exp_{\mathbf{X}_a|\mathbf{X}_e} \left[\mathcal{Y}(\mathbf{X},a)\right] ,a)\right] 
\end{equation}
A sufficient condition for Eq. \ref{eq:linear_utility_Xa} to hold is that $u'$ is linear in $\mathcal{Y}$. If Eq. \ref{eq:linear_utility_Xa} does not hold, the QoI must be defined as the utility function itself, at least for the purpose of the sensitivity analysis. 


Including a first integration step and working only with the epistemic uncertainty, as in Eqs. \ref{eq:utility_function_Epistemic} and \ref{eq:general_model_epistemic}, can have computational advantages over working with all $\mathbf{X}$ explicitly.
This is due to the reduced variance of the model outcome $Y_e$ in Eq. \ref{eq:general_model_epistemic} compared to $Y$ in Eq. \ref{eq:general_model}. 
However, this is only beneficial if the evaluation of expectation in Eq. \ref{eq:general_model_epistemic} is computationally cheap.

\section{Working example}
\label{sec:WorkingExample}

To illustrate the concepts presented in the following sections, we utilize a simple example, which is introduced in the following. 
Consider three different alternatives $a \in \{1,2,3\}$ for a risk protection system. The QoI
 $Y$ is the losses (damages) associated with the choice $a$ over the considered time-span, which is 
\begin{equation} \label{eq:WorkingExampleLosses}
	Y(a) = \mathcal{Y}(\mathbf{X},a)=
	\begin{cases}
		0, & S < R_a, \\
		C_F \cdot \left(S - R_a\right), & S \ge R_a, 
	\end{cases}
\end{equation}
where $S$ is the maximum demand (load) on the system, $R_a$ is the capacity (resistance) of the protection system $a$ and $C_F$ is a factor determining the costs of exceeding the capacity of the protection system. The model parameters are summarized in Table \ref{tab:prob_model_example1}. 

\begin{table}[]
	\centering
	\caption{Parameters of the working example. $\gamma=0.57721$ denotes Euler's constant.}
	\label{tab:prob_model_example1}
	\small{
	\begin{tabular}{lllll}
		\hline
		 Parameter & Distribution & Mean & standard dev. \\ \hline
        Load $S$    & Gumbel & $M + \gamma$ & $\pi / \sqrt{6} \approx 1.2825$ \\
		Gumbel parameter $M$  & normal& $7.5$ & $1$ \\
        Resistance $R_1$          & lognormal       &    $10$  &  $1$                    \\
        Resistance $R_2$          & lognormal       &    $12$  &  $1$                    \\
        Resistance $R_3$          & lognormal       &    $14$  &  $1$                    \\
		Cost of failure $C_F$ [\euro] & lognormal& $3\cdot 10^7$ & $10^7$           \\                   \hline
	\end{tabular}
}
\end{table}

The load $S$ is Gumbel distributed with scale parameter $\beta=1$ and location parameter $M$, which is itself an uncertain parameter. 
The probability density function (PDF) of $S$ is 
\begin{equation}\label{eq:Gumbel_PDF}
 	f_S(s;M) = \exp \left[ -s+M -\exp \left(-s+M\right) \right],
\end{equation}
where the notation $f_S(s;M)$ is used to make it explicit that the PDF of $S$ depends on the parameter $M$.

In this example, the load is the only aleatory uncertainty, i.e., $\mathbf{X}_a=\left[S\right]$. The epistemic uncertainty is characterized by the remaining random variables, i.e., $\mathbf{X}_e=\left[M;R_1;R_2;R_3;C_F\right]$, which includes the statistical uncertainty on $S$ represented through $M$. 

Each potential protection system is associated with a cost $c_a$, which is reported in Table \ref{tab:costs_example1}. 
One wants to select the system with the lowest sum of expected cost and damages. Hence the utility function is
\begin{equation} \label{eq:utility_working_example_original}
    u(\mathbf{X},a) = -\mathcal{Y}(\mathbf{X},a)-c_a(a).
\end{equation}

Following Section \ref{sec:alea_epistemic}, there is no point in determining the sensitivity with respect to the aleatory uncertainty factor $S$.  Therefore, the quantity of interest is the expected value of $Y(a)$ with respect to this factor $S$, 
\begin{equation}\label{eq:WorkingExampleExpectedValue}	Y_e(a) = \mathcal{Y}_e(\mathbf{X}_e,a) = \Exp_S\left[\mathcal{Y}([S,\mathbf{X}_e],a)\right]=\int_{R_a}^{\infty}C_F(s-R_a)f_S(s;M)\,\text{d}s.
\end{equation}



Since the utility function of Eq. \ref{eq:utility_working_example_original} is linear with respect to $Y(a)$, the analysis can be performed with the following utility function instead:
\begin{equation} \label{eq:utility_working_example}
    u(\mathbf{X}_e,a) = -\mathcal{Y}_e(\mathbf{X}_e,a)-c_a(a).
\end{equation}

The resulting expected values for all options are summarized in Table \ref{tab:costs_example1}.
The preferred system is $a_{opt}=2$. 

\begin{table}[]
	\centering
	\caption{Costs and expected losses associated with the three protection systems.}
	\label{tab:costs_example1}
	\small{
	\begin{tabular}{lllll}
		\hline
		 Protection  & Cost $c_a$ & Expected loss & Expected utility [$10^6$\euro] \\system $a$ &  [$10^6$\euro] & [$10^6$\euro] & (negative expected sum of cost and losses)\\ \hline
        $1$    & $13$ & $5.31$ & $-18.31$ \\
        $2$ & $15$ & $0.84$ & $-15.84$	\\
        $3$ & $17$ & $0.12$ & $-17.12$ \\
        \hline
	\end{tabular}
}
\end{table}



\begin{figure}[h] \centering
    \includegraphics[width=1\textwidth]{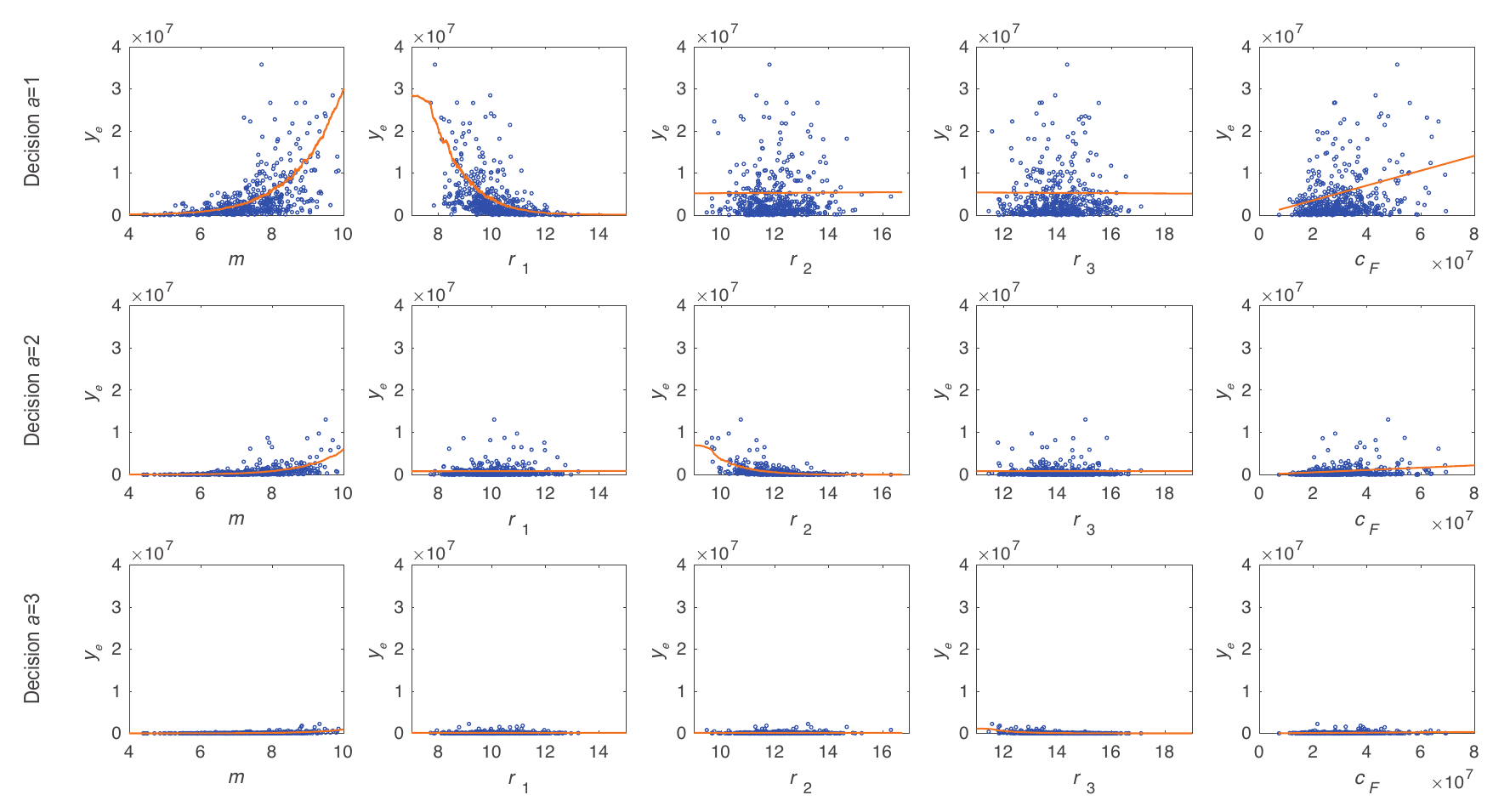}
    \caption{Scatter plot of input variables against the output $Y(a)$ according to Eq. \ref{eq:WorkingExampleExpectedValue}. The three rows represent the different decision alternatives $a=1,2,3$ The orange lines show estimates of the conditional expectation of the losses given the marginal inputs.}
    \label{fig:Fig_Scatter_matrix}
\end{figure}

Figure \ref{fig:Fig_Scatter_matrix} shows scatter plots of $Y_e$ against the inputs $\mathbf{X}_e=\left[M,R_1,R_2,R_3,C_F\right]$, generated with $500$ Monte Carlo samples.
The results are shown in function of the decision. In Figure \ref{fig:Fig_Scatter_matrix}, one can see that the losses are only sensitive to $R_1$ if the decision is $a=1$, but not if $a=2$ or $a=3$, whereas they are only sensitivie to $R_2$ if the decision is $a=2$, and so on. Additionally, for decisions $a=2$ and, in particular, $a=3$, the resulting losses are less sensitive to the inputs in an absolute sense. The scale on the vertical axes is purposely kept the same in all three rows of Figure \ref{fig:Fig_Scatter_matrix}. If we adjust the scale of the axes in the second and third rows, the images in those rows would be similar to those in the first row. 

\begin{figure}[h] \centering
    \includegraphics[width=0.7\textwidth]{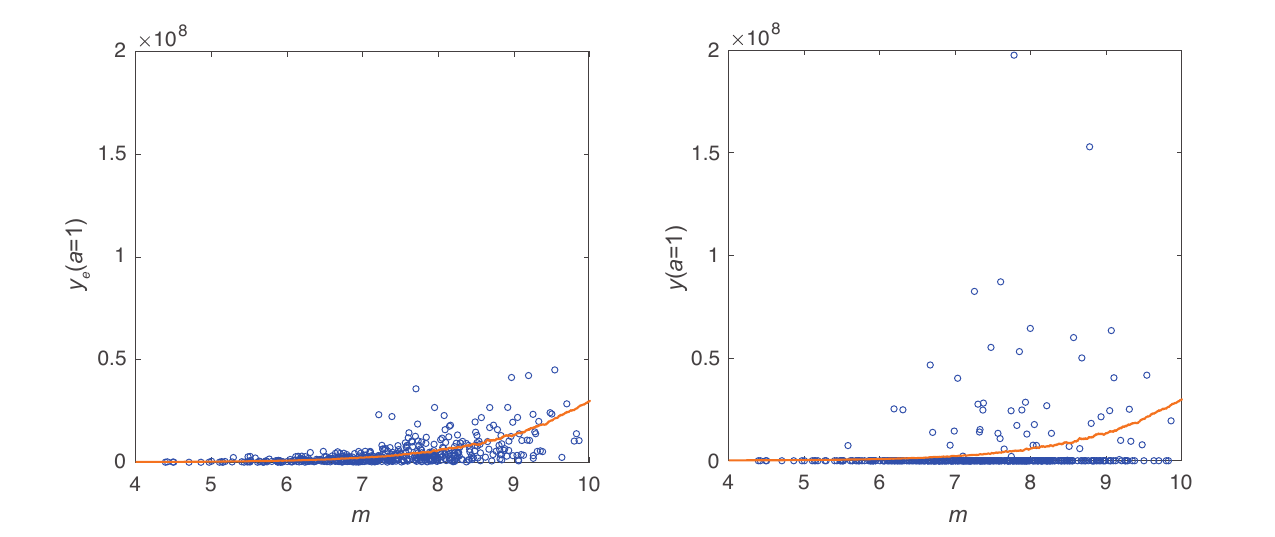}
    \caption{Scatter plot of factor $M$ against the QoI. Left-hand side: The QoI is $Y_e$, the expectation of the loss with respect to $S$ following Eq.~\ref{eq:WorkingExampleExpectedValue}. Right-hand side: The QoI is $Y$, i.e., the loss in function of $S$, following Eq.~\ref{eq:WorkingExampleLosses}. The orange lines show estimates of the conditional expectation $\Exp[Y|M=m]$.}
    \label{fig:Fig_Scatter_Epistemic_vs_aleatory}
\end{figure}

To illustrate the effect of eliminating aleatory uncertainty in the analysis, Figure \ref{fig:Fig_Scatter_Epistemic_vs_aleatory} compares the scatter diagrams of $M$ against $Y_e(a=1)$ with the one of $M$ against $Y(a=1)$.
The left-hand side is a copy of the upper left panel in Figure \ref{fig:Fig_Scatter_matrix}, where the axis is rescaled to match the one on the right. 
In the Monte Carlo samples shown on the right-hand side, $S$ is handled by sampling as well. The scatter in the resulting model outcomes is much larger, and the effect of $M$ on the losses is less obvious, even if the smoothing function $\Exp[Y(a=1)|M]$ is the same as the function $\Exp[Y_e(a=1)|M]$ on the left-hand side. 

\section{Decision-theoretic sensitivity measures}
\label{sec:decision_sensitivity}


\subsection{Information value as a sensitivity measure}
\label{sec:Information_value}

To derive decision sensitivity measures for factor prioritization, one quantifies how perfectly knowing the input $X_i$ influences the expected utility. If it is known that $X_i$ takes the value $x_i$, the optimal decision becomes\footnote{Here and in the following, we describe the theory for the case of the full model $\mathcal{Y}(\mathbf{X},a)$ but note that it is equally valid when working with the reduced model $\mathcal{Y}_e(\mathbf{X}_e,a)$.}

\begin{equation}
	\label{eq:posterior_decision}
	a_{opt|X_i}(x_i)=\arg \max_a \Exp_{\mathbf{X}|X_i=x_i}\left[u(\mathbf{X},a)\right],
\end{equation}
wherein $\Exp_{\mathbf{X}|X_i=x_i}$ is the expectation with respect to $\mathbf{X}$ conditional on $\{X_i=x_i\}$. 

Under the observation $\{X_i=x_i\}$, the a-priori optimal decision $a_{opt}$ of Eq. \ref{eq:prior_decision} may no longer be optimal. The expected utility gain of changing the decision from $a_{opt}$ to $a_{opt|X_i}(x_i)$ in this case is 
\begin{equation} \label{eq:CVPPI}
CVPPI_{X_i}(x_i)=\Exp_{\mathbf{X}|X_i=x_i}\left[u(\mathbf{X},a_{opt|X_i}(x_i))\right]-\Exp_{\mathbf{X}|X_i=x_i}\left[u(\mathbf{X},a_{opt})\right].
\end{equation}
$CVPPI_{X_i}$ stands for Conditional Value of Partial Perfect Information associated with input $X_i$. In the following, we drop the argument $x_i$ in the expression for the conditionally optimal decision $a_{opt|X_i}$ to simplify notation.

The Expected Value of Partial Perfect Information (EVPPI) for $X_i$, in short \textit{Information Value} $V_{X_i}$, is the expectation of the $CVPPI_{X_i}$ with respect to $X_i$:
\begin{equation}
	\label{eq:IV}
	\begin{split}
		V_{X_i}&= \Exp_{X_i}\left[CVPPI_{X_i}(X_i)\right] \\ 
		&= \Exp_\mathbf{X}\left[u(\mathbf{X},a_{opt|X_i})\right]-\Exp_{\mathbf{X}}\left[u(\mathbf{X},a_{opt})\right]
	\end{split}
\end{equation}


The information value $V_{X_i}$  quantifies the expected gain in utility achieved by full knowledge of $X_i$. It provides an upper bound on the benefit one can achieve by learning about $X_i$. This is an upper bound on the expected gain, because imperfect (partial) information on $X_i$ cannot have a larger value \cite{Raiffa1961}, see also Section \ref{sec:sample_value}.

The information value can also be computed for a group $\upsilon$ of input variables $\mathbf{X}_{\upsilon} = \{X_i,\,i \in \upsilon\}$. In this case, it is 
\begin{equation}
	\label{eq:IV_group}            
    V_{\mathbf{X}_{\upsilon}} = \Exp_\mathbf{X}\left[u(\mathbf{X},a_{opt|\mathbf{X}_{\upsilon}})\right]-\Exp_{\mathbf{X}}\left[u(\mathbf{X},a_{opt})\right]. 
\end{equation}

Note that in general the $V_{\mathbf{X}_{\upsilon}}$ is not equal to the sum of the individual information values $V_{X_i}$ of all $i \in \upsilon$. 
Indeed, the information value can be subadditive, i.e., $V_{\mathbf{X}_{\upsilon}} < \sum_{\,i \in \upsilon} V_{X_i}$, or superadditive, i.e., $V_{\mathbf{X}_{\upsilon}} > \sum_{\,i \in \upsilon} V_{X_i}$ \cite{samson_value_1989,frazier_paradoxes_2010}.

The above equations are valid when working with all random variables $\mathbf{X}$ and the QoI $Y$ or when working only with the epistemic uncertainty factors $\mathbf{X}_e$ and the QoI $Y_e$. The resulting information value for the factors in $\mathbf{X}_e$ will be the same, independent of the QoI used. The difference between the two QoI lies in the accuracy of sampling-based approximations of the information values due to the larger variance of $Y$ compared to $Y_e$.

\subsection{Relative and normalized information value} \label{sec:normalize}
In some instances, it is desirable to normalize the information value. Following \cite{straub2022decision}, a natural way to normalize it is by means of the Expected Value of Perfect Information (EVPI), i.e., the expected gain in utility by achieving perfect knowledge on all uncertain inputs $\mathbf{X}$, which is
 \begin{equation}
 	\label{eq:EVPI}
 		EVPI=  \Exp_\mathbf{X}\left[u(\mathbf{X},a_{opt|\mathbf{X}})\right]-\Exp_{\mathbf{X}}\left[u(\mathbf{X},a_{opt})\right]. 
 \end{equation}
wherein $a_{opt|\mathbf{X}}$ indicates the optimal decision under full knowledge of all uncertain inputs, i.e., the result of a decision making under certainty. 

However, considering that it is impossible to learn about the inherent randomness (aleatory uncertainty), a fairer way to normalize is by means of the Expected Value of the Perfect Model, wherein  all epistemic uncertainty is assumed to be known when making the decision: 
 \begin{equation}
	\label{eq:EVPM}
	EVPM=  \Exp_\mathbf{X}\left[u(\mathbf{X},a_{opt|\mathbf{X}_e})\right]-\Exp_{\mathbf{X}}\left[u(\mathbf{X},a_{opt})\right]. 
\end{equation}
In general the EVPM is more challenging to compute than the EVPI, as discussed in Section \ref{sec:Computing}. If there is no aleatory uncertainty, or if the quantity of interest is the expected value with respect to the aleatory uncertainty $Y_e$, then the EVPM is equal to the EVPI.

We refer to the information values normalized with EVPI or EVPM as relative information values:
\begin{equation}
	\label{eq:relativeIV}
	\text{relative }V_{X_i}= \frac{V_{X_i}}{EVPI} \quad \text{or} \quad \text{relative }V_{X_i}= \frac{V_{X_i}}{EVPM} 
\end{equation}



\subsection{Probability of decision change as a sensitivity measure}
As an alternative to the information value, it is also possible to consider the probability of a decision change as a sensitivity measure, for cases with discrete decision alternatives. That is, we define the sensitivity measure
\begin{equation}
	\label{eq:PrDecChange}
		DC_{X_i}= \Pr\left(a_{opt|X_i}\ne a_{opt}\right).
\end{equation}
This sensitivity measure is readily obtained as a side-product when computing $V_{X_i}$. 

\subsection{Working example}
Table \ref{tab:IV_example1} summarizes the information value of the different inputs of the working example. The table does not show the information value of $S$, which can nevertheless be computed as well. It is $V_S=1.88\cdot 10^6$\euro. The reason for not reporting it in Table \ref{tab:IV_example1} is that $S$ is an aleatory uncertainty and there is nothing that can be done to reduce it for a given value of $M$.

The expected value of the perfect model is $EVPM=0.92\cdot 10^6$\euro, which is used to calculate the relative information value in Table \ref{tab:IV_example1}. When working with the QoI $Y_e$ according to Eq. \ref{eq:WorkingExampleExpectedValue}, this is also the EVPI. However, when the QoI is the original $Y$ of Eq. \ref{eq:WorkingExampleLosses} in which $S$ is an input factor, the EVPI is significantly larger, $EVPI=2.3\cdot 10^6$\euro. This should not be surprising: If the maximum load $S$ on the system were known in advance, the choice of the optimal level of resistance becomes much easier. One would simply choose the lowest system capacity larger than $S$. However, since it is impossible to know $S$, this EVPI can be misleading and it is preferable to communicate the EVPM.

\begin{table}[]
	\centering
	\caption{Sensitivity indices of the working example.}
	\label{tab:IV_example1}
	\small{
	\begin{tabular}{lllll}
		\hline
		 Parameter & Information value & Relative information & Probability of decision \\ & $V_{X_i}$ [$10^3$\euro] &  value $V_{X_i}/EVPM$ & change $DC_{X_i}$\\ \hline
		Gumbel parameter $M$  & $336$ & $36.5\%$ & $0.37$\\
        Resistance $R_1$     &    $459$  &  $49.8\%$ & $0.36$                   \\
        Resistance $R_2$     &    $81$  &  $8.8\%$    &     $0.07$           \\
        Resistance $R_3$    &    $0$  &  $0$     &     $0$          \\
		Cost of failure $C_F$ & $2$ & $0.3\%$   &       $0.01$ \\                   \hline
	\end{tabular}
}
\end{table}

\begin{figure}[h] \centering
    \includegraphics[width=0.7\textwidth]{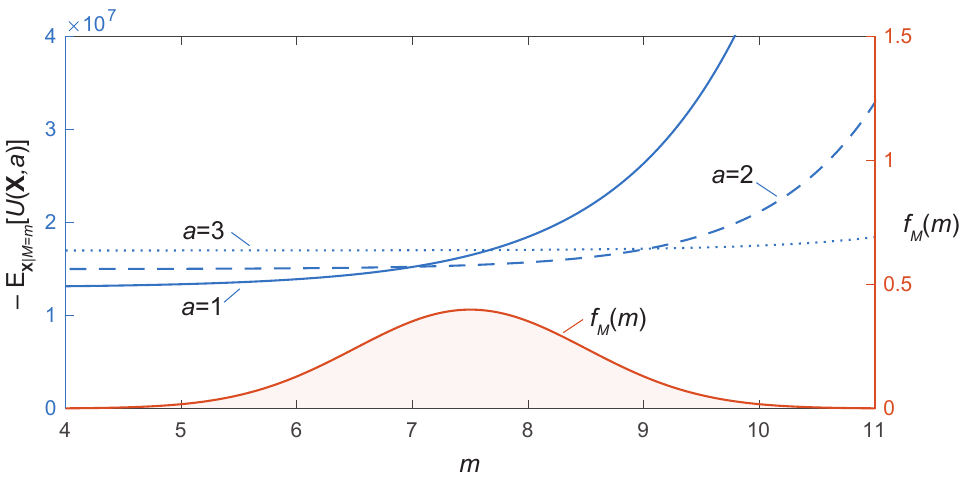}
    \caption{Expected utility conditional on $M$ and $a$, $\Exp_{\mathbf{X}|M=m}\left[u(\mathbf{X},a)\right]$, together with the PDF of $M$. }
    \label{fig:Fig_CondU}
\end{figure}

To better understand the sensitivity to selected parameters, additional plots can be useful, in particular a comparison of the conditional utility with different decision alternatives. Exemplary, Figure \ref{fig:Fig_CondU} shows the expected utilities conditional on $M$, $\Exp_{\mathbf{X}|M=m}\left[u(\mathbf{X},a)\right]$. Note that the difference between the conditional utility obtained with the a-priori choice $a_{opt}=2$ and the optimal choice for a given $M=m$ is the CVPPI, see Eq. \ref{eq:CVPPI}. For values of $m$ smaller than approx. $7$, the conditionally optimal decision is $a_{opt|M}=1$ and for values of $m$ larger than approx. $9$, the conditionally optimal decision is $a_{opt|M}=3$. 

\subsection{Sample information value}\label{sec:sample_value}
In some applications, one might have a good idea on how much can actually be learnt on a specific input quantity. For example, if $X_i$ is subject to statistical uncertainty, one might specify the amount of additional data (e.g., the number of samples) that can be collected. 
In these cases, one might wish to quantify the information value associated with this data, rather than the value of perfectly knowing $X_i$. For this reason, Oakley \cite{Oakley2008} suggests the use of the expected value of sample information as a sensitivity measure. To this end, the data that can potentially be obtained on $X_i$ is quantified by the likelihood $L_i$ and the effect of the data on $X_i$ and $\mathcal{Y}(\mathbf{X})$ is evaluated by means of Bayesian analysis. 

Oakley \cite{Oakley2008} proposes the use of a Gaussian process regression together with conjugate likelihoods for computing the sampling information value. However, we prefer the more flexible and conceptually simpler approach proposed by Strong et al. \cite{strong2015estimating}. To this end, we note that -- a-priori -- one can consider the data or their sufficient statistics as a random vector $\mathbf{Z}$ that is dependent on $\mathbf{X}$ via the likelihood function $L(\mathbf{x}|\mathbf{z})$:
\begin{equation} \label{eq:z_cond_PDF}
	f_{\mathbf{Z}|\mathbf{X}}(\mathbf{z}|\mathbf{x})=L(\mathbf{x}|\mathbf{z})
\end{equation}
Conceptually, one can obtain the sample information value by performing the sensitivity analysis with an augmented input vector $\mathbf{X}'=[\mathbf{X};\mathbf{Z}]$ and then evaluating the information value of $\mathbf{Z}$. The joint PDF of $\mathbf{X}'$ is simply $f_{\mathbf{X}'}(\mathbf{x}') = f_{\mathbf{Z}|\mathbf{X}}(\mathbf{z}|\mathbf{x}) f_{\mathbf{X}}(\mathbf{x})$

Following Eq. \ref{eq:IV_group}, the information value of $\mathbf{Z}$, i.e., the sample information value, is
\begin{equation}
	\label{eq:sample_value}
		V_{\mathbf{Z}}=  \Exp_{\mathbf{X'}}\left[u(\mathbf{X},a_{opt|\mathbf{Z}})\right]-\Exp_{\mathbf{X}}\left[u(\mathbf{X},a_{opt})\right]. 
\end{equation}

This concept can be used more generally to evaluate the information value of any type of data collection, including that from performing a test or installing a sensor, as long as the likelihood function can be specified. Examples of such value-of-information analyses can be found in the literature \cite{straub_value_2014,malings_value_2016,thons_value_2018,kamariotis_framework_2023}, which use dedicated algorithms for evaluating the sample information value.
However, as we show in Section \ref{sec:Computing}, the sample information value can be computed cheaply within a Monte Carlo framework. 

To illustrate the relation between the sample information value and the information value, we reconsider the working example and assume that one can reduce the uncertainty in parameter $M$ by collecting additional samples $\mathbf{S} = S_1,\dots,S_{n_s}$ of the demand $S$. The likelihood function describing this data follows from Eq. \ref{eq:Gumbel_PDF} as
\begin{equation}\label{eq:likelihood_example}
	L(m|\mathbf{s})=\prod_{i=1}^{n_s} f_S(s_i;m) = \exp \left[ \sum_{i=1}^{n_s}  -s_i+m -\exp \left(-s_i+m\right) \right].
\end{equation}
For this likelihood, a sufficient statistics of the data is 
\begin{equation}\label{eq:sufficientStatistics}
	Z= \sum_{i=1}^{n_s}  \exp (-S_i).
\end{equation}
We sample $Z$ by first sampling $S_1,\dots,S_{n_s}$ conditional on the sample of $M$ and then obtain the corresponding $Z$ sample via Eq. \ref{eq:sufficientStatistics}. Then, we use the $Z$ samples together with the corresponding $\mathbf{X}$ samples to determine the sample information value $V_Z$ following Eq. \ref{eq:sample_value}.

\begin{figure}[h] \centering
    \includegraphics[width=0.8\textwidth]{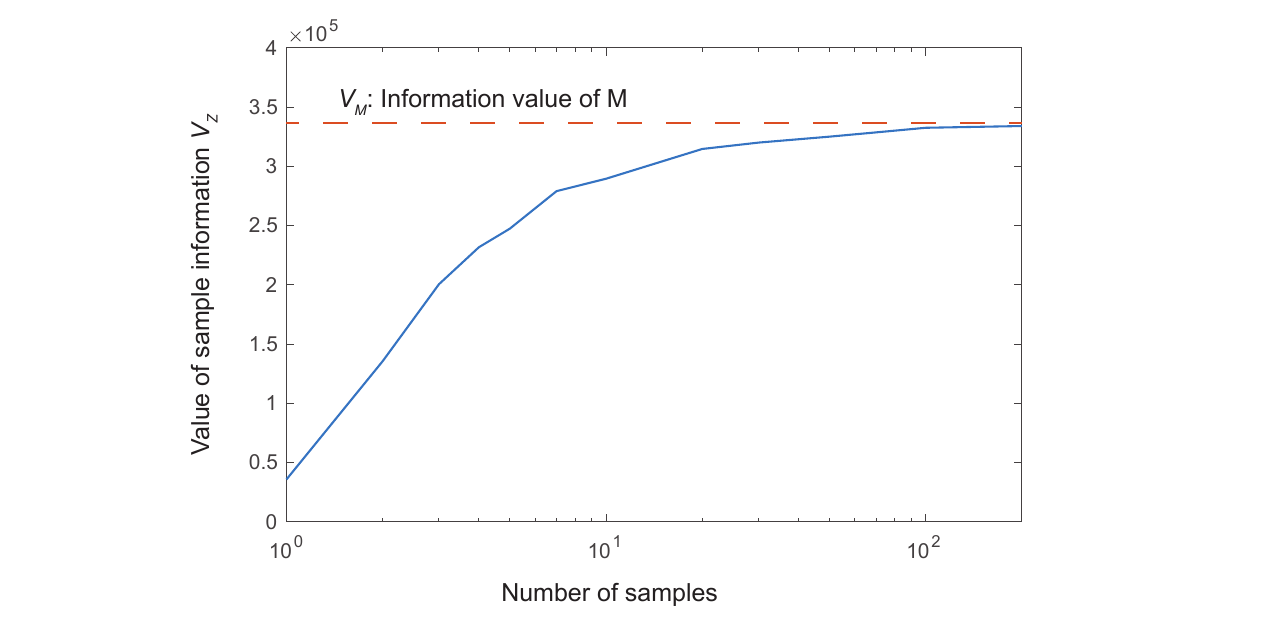}
    \caption{Value of sample information $V_Z$ for varying number $n_s$ of load samples $S_i, \, i=1,\dots,n_s$. }
    \label{fig:Fig_Value_of_Sample_Information}
\end{figure}

Figure \ref{fig:Fig_Value_of_Sample_Information} shows the sample information value in function of the number of collected samples $n_s$. With increasing $n_s$, the value of sample information asymptotically approaches the information value of $M$. This is expected because the samples enable learning of $M$ only. Hence, perfect information on $M$ must be an upper bound to $V_Z$ in this case. 

\section{Modeling decisions and utility} \label{sec:decisions_utility}
The derivation of the information value requires the definition of the decision alternatives and the utility function. We suggest to distinguish between the following types of decisions:
\begin{enumerate}
	\item binary decisions, where the choice is between two alternative actions;
	\item discrete decisions, where the choice is among a discrete set of alternative actions;
	\item continuous decision, where one chooses one or more continuous parameters;
	\item sequential decision settings, where multiple decisions are taken in a sequential manner, possibly informed by observations. These decisions can be binary, discrete, or continuous.
\end{enumerate}
We discuss discrete and continuous decisions in the two subsequent sections. The binary decision case is a special case of the discrete decision case and we do not treat it separately in the following. Furthermore, we do not discuss sequential decision settings here and leave this for future work. We only note that in such cases it is often possible to identify a simplified proxy decision setting for the purpose of the sensitivity analysis. As an example, when considering the use of a prognostics model for maintenance planning \cite[e.g.,]{kamariotis_metric_2024}, one can investigate a decision on repair or replacement at a single time instance $t$.

\subsection{Discrete decisions}
With discrete decisions, the choice is among a finite set of alternatives $a \in \{1,2,\dots,n_a\}$.
The most basic case is the binary decision case, with $n_a=2$. Examples include the decision between retrofitting an existing engineering system and doing nothing, or the decision between implementing and not implementing a system.
Multiple decision alternatives arise, e.g., as a decision among multiple options for an infrastructure project, or a decision among multiple solutions for an engineering problem as in the working example.


The utility function must be defined case-specific. In many applications, the utility function will be equal to the benefits minus cost. For example, in risk-based decision-making, a typical utility function is
\begin{equation} \label{eq:u_risk}
    u(\mathbf{X}_e,a) = - p_F(\mathbf{X}_e,a)c_F -c_a(a),
\end{equation}
where $p_F$ is the probability of failure in function of $\mathbf{X}_e$ and $a$, $c_F$ is the cost of a failure and $c_a$ is the cost of the choice $a$. 
Note that the probability of failure is an expected value with respect to aleatory uncertainties. 

While it can be convenient to express utility in monetary terms, this is not always possible or desirable. In the assessment of erosion at sites for a deep geological repository, presented in Section \ref{sec:application_nuclearWaste}, the choice is not based on monetary considerations. Instead, the utility is based on the probability of exposure alone, i.e., it is simply 
\begin{equation} \label{eq:u_pF}
    u(\mathbf{X}_e,a) = - p_F(\mathbf{X}_e,a).
\end{equation}


\subsection{Continuous decisions} \label{sec:continuous_decisions}
Continuous decisions arise typically when designing a system, e.g., when choosing the capacity of a new infrastructure or the dimensions of an engineering structure. In general, one can have multiple design parameters, e.g., the capacity/dimension of system components. However, for the purpose of sensitivity analysis, such a model will typically be unnecessarily complex, and we consider only the case of a single design parameter. 

We distinguish the cases in which the parameter $a$ is part of the model $\mathcal{Y}$ from those in which it is not. In the latter cases, the utility function of Eq. \ref{eq:define_utility_function_Y} reduces to $u(\mathbf{X},a)=u'\left(\mathcal{Y}(\mathbf{X}),a\right)$. Such utility functions can arise when the decision is on the capacity $a$ of a system and $Y=\mathcal{Y}(\mathbf{X})$ is the uncertain demand. An important instance of such a utility function is 
\begin{equation}
	\label{eq:u_quadratic}
	u(\mathbf{X},a) = - c \left(\mathcal{Y}(\mathbf{X})-a\right)^2,
\end{equation}
in which $c$ is a positive constant. 
This function implies that the utility decreases with the square of the deviation of the system capacity from the system demand. 

Another instance of a utility function $u(\mathbf{X},a)=u'\left(\mathcal{Y}(\mathbf{X}),a\right)$ is the LINEX (linear exponential) function \cite{Varian1975,Zellner1986}, which can represent asymmetric consequences: 
\begin{equation}
	\label{eq:u_LINEX}
	u(\mathbf{X},a)=-c \left\{ \exp\left\{\gamma\left[\mathcal{Y}(\mathbf{X})-a\right]\right\}-\gamma\left[\mathcal{Y}(\mathbf{X})-a\right]-1 \right\}
\end{equation}
The parameter $\gamma >0$ controls the degree of asymmetry, with $\gamma \to 0$ approaching symmetry and increasing values of $\gamma$ corresponding to higher asymmetry. 

\begin{figure}[h] \centering
    \includegraphics[width=0.7\textwidth]{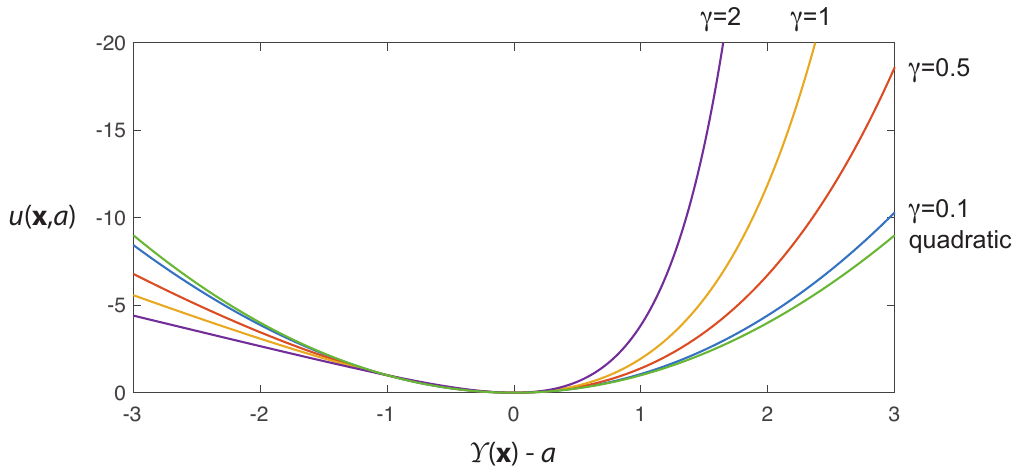}
    \caption{Quadratic and LINEX utility functions with varying $\gamma$. The parameter $c$ of Eqs. \ref{eq:u_quadratic} and \ref{eq:u_LINEX} is chosen such that the utility function results in $1$ for $(a-\mathcal{Y})=-1$.}
    \label{fig:LINEX_utility}
\end{figure}

LINEX represents situations in which the losses associated with the capacity $a$ not meeting the demand ($a<\mathcal{Y}(\mathbf{X})$) are higher than the losses associated with an over-design of the system ($a>\mathcal{Y}(\mathbf{X})$). That is typical for the design of infrastructure systems, where insufficient capacity can lead to costs that are significantly higher than the costs for over-design. 
The quadratic and LINEX utility functions are illustrated in Figure \ref{fig:LINEX_utility}.

More generally, $a$ can be parameters of the model $\mathcal{Y}(\mathbf{X},a)$. For example, $a$ can be a design parameter of an engineering system or a control parameter in an environmental system. In these cases, the utility function will be similar to Eq. \ref{eq:u_risk}, with the system model describing the performance (e.g., the probability of failure, or the system output) in function of $\mathbf{X}$ and $a$. 

\textit{Remark}. With the quadratic utility function of Eq. \ref{eq:u_quadratic}, the relative information value reduces to the first-order Sobol' index \cite{oakley2004probabilistic,Borgonovo2021}. This can be seen easily, as the optimal decision with this utility function is $a_{opt}=\Exp\left[\mathcal{Y}(\mathbf{X})\right]$ and the conditional value of partial perfect information of Eq. \ref{eq:CVPPI} becomes the conditional variance $\text{Var}\left[\mathcal{Y}(\mathbf{X})|X_i=x_i\right]$. Because the EVPI with the quadratic loss function is $\text{Var}\left[\mathcal{Y}(\mathbf{X})\right]$, the Sobol' index is equal to the relative information value. 
This result signifies that the Sobol' index is a suitable sensitivity measure when decisions are reflected by the quadratic utility function of Eq. \ref{eq:u_quadratic}, but might not be a good measure otherwise. Furthermore, because the LINEX utility function reduces to the quadratic utility function for $\gamma \to 0$, the information value with the LINEX utility function can be interpreted as a generalization of the Sobol' index.

\subsection{Working example} \label{sec:working_example_continuous}

To illustrate the case of continuous decisions, we alter the working examples as follows: Instead of the discrete choice of an option, a decision $a$ can now be taken on the mean value of the resistance, which can be selected as any positive real number. The resistance $R$ is the product of $a$ with a lognormal distributed model uncertainty $X_R$ with mean $1$ and standard deviation $0.1$. 
The cost associated with design $a$ is $c_a(a)=a\cdot10^6 + 3\cdot 10^6$. The remaining parameters of the problem are unaltered. Therefore, the QoI of the continuous problem is
\begin{equation}\label{eq:WorkingExampleExpectedValue_continuous} \mathcal{Y}_e(\mathbf{X}_e,a) =\int_{a\cdot X_R}^{\infty}C_F(s-a\cdot X_R)f_S(s;M)\,\text{d}s,
\end{equation}
by analogy with Eq. \ref{eq:WorkingExampleExpectedValue}.
The utility function remains as in Eq. \ref{eq:utility_working_example}
.

\begin{figure}[h] \centering    \includegraphics[width=0.6\textwidth]{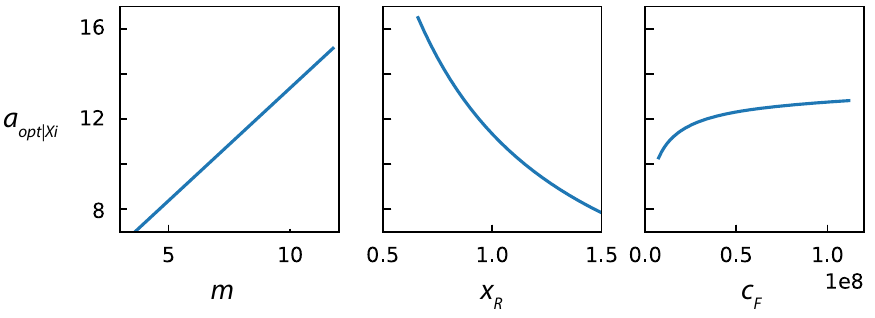}
    \caption{Conditionally optimal $a_{opt|X_i}$ for the continuous version of the working example.}
    \label{fig:cond_aopt_continuous}
\end{figure}

The optimal a-priori decision is $a_{opt}=11.9$, with associated expected total costs and damages of $\Exp[Y(a_{opt})]=15.98\cdot10^6$. Figure \ref{fig:cond_aopt_continuous} shows the conditionally optimal decisions $a_{opt|X_i}$. The resulting information values are summarized in Table \ref{tab:IV_example_continuous}. The EVPM is evaluated following Section \ref{sec:EVPM_calculation} as  $EVPM=1.04\cdot10^6$ and is used to compute the relative information values, which are also reported in Table \ref{tab:IV_example_continuous}.

\begin{table}[]
	\centering
	\caption{Sensitivity indices of the continuous version of the working example.}
	\label{tab:IV_example_continuous}
	\small{
	\begin{tabular}{lllll}
		\hline
		 Parameter & Information value & Relative information & Sobol' index \\ & $V_{X_i}$ [$10^3$\euro] &  value $V_{X_i}/EVPM$ & \\ \hline
        		Gumbel parameter $M$  & $404$ & $39\% $ & $26\% $\\
                Resistance uncertainty $X_R$     &    $492$  &  $47\%$ & $ 30\% $                   \\
		Cost of failure $C_F$ & $61$ & $6\%$   &       $2\% $ \\ 
                  \hline
	\end{tabular}
}
\end{table}

For comparison, we also evaluate the first-order Sobol' index for the model in which $a$ is set to the a-priori optimal value $a_{opt}$. 
The results are also reported in Table \ref{tab:IV_example_continuous}. We observe that the Sobol' indices give the same ranking as the relative IVs but are consistently lower.








\section{Sampling-based computation of decision sensitivity measures}
\label{sec:Computing}

We assume that a sampling-based uncertainty quantification is performed, i.e., the statistics of $Y(a)=\mathcal{Y}(\mathbf{X},a)$ are determined based on a set of (possibly weighted) samples of $Y$. Sampling-based approaches include Monte-Carlo methods, low-discrepancy sequences, and importance sampling\footnote{We do not explicitly discuss the treatment of weighted samples that result from importance sampling. The expressions given here assume unweighted (or uniformly weighted) samples and have to be adjusted to consider important sampling weights. Alternatively, one can perform first a resampling step to obtain uniformly weighted samples and then apply the provided equations. } \cite{rubinstein2016simulation,mcbook}.
In the following, we present methods to compute the decision sensitivity indices based on post-processing these samples. In general, the computationally expensive part of the analysis is the evaluation of the model $\mathcal{Y}(\mathbf{X},a)$. Therefore, methods based on post-processing these samples, without requiring additional model evaluations, can be considered computationally cheap. 

To determine the decision sensitivity, one must (a) evaluate the (conditional) expectation of the utility function $\Exp_{\mathbf{X}}\left[u(\mathbf{X},a)\right]$ and $\Exp_{\mathbf{X}|\mathbf{X}_{\upsilon}=\mathbf{x}_{\upsilon}}\left[u(\mathbf{X},a)\right]$ and (b) solve the optimization problems in Eqs. \ref {eq:prior_decision} and \ref{eq:posterior_decision}. In the following, we discuss special (but fairly common) cases in which these quantities can be evaluated with limited computation efforts. 


\subsection{Discrete decision alternatives}
We first consider the case of discrete decision alternatives $a \in \{1,\dots,n_a\}$. In this case, one must evaluate the conditional expected utility $\Exp_{\mathbf{X}|X_{\upsilon}=\mathbf{x}_{\upsilon}}\left[u(\mathbf{X},a)\right]$ for all $n_a$ decision alternatives separately. The optimization is then trivial. We assume that samples of $Y(a) = \mathcal{Y}(\mathbf{X},a)$ are available for all $a$. 

$\Exp_{\mathbf{X}|\mathbf{X}_{\upsilon}=\mathbf{x}_{\upsilon}}\left[u(\mathbf{X},a)\right]$ is approximated using samples of $u(\mathbf{X},a)=u'(Y(a),a)$. These samples are obtained with negligible computational effort from samples of $Y(a)$, because the utility function is cheap to evaluate (it is typically of an analytical form). 

\begin{figure}[h] \centering
    \includegraphics[width=1\textwidth]{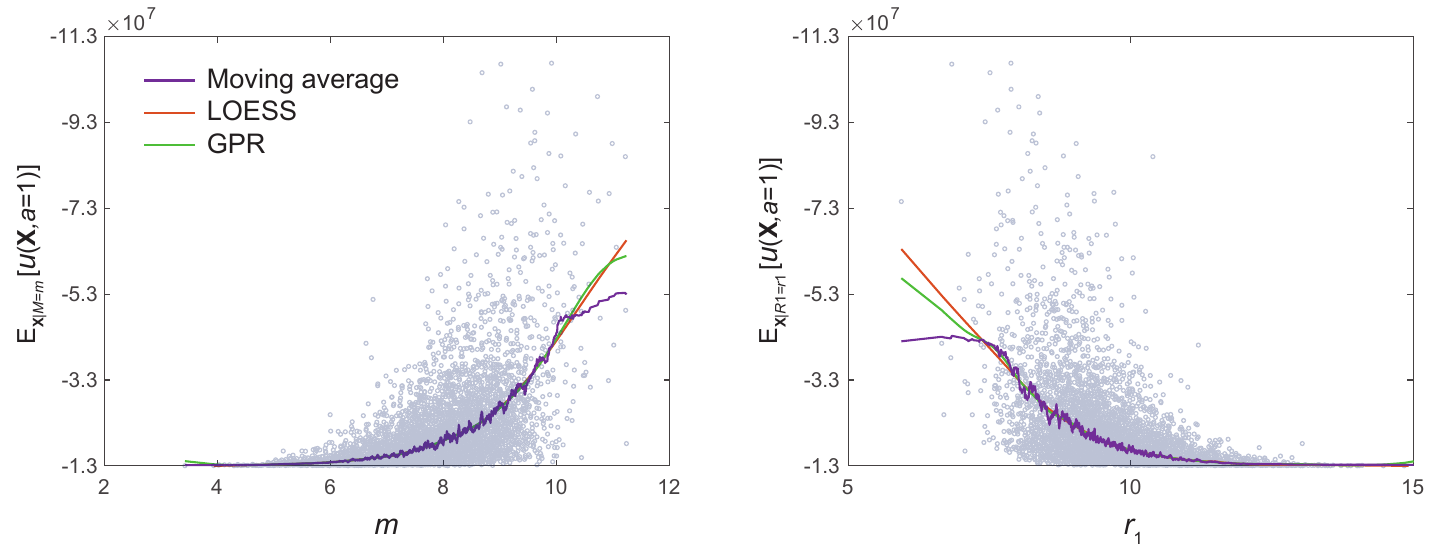}
    \caption{Conditional expectation of the utility function given inputs $M$ (left panel) and $R_1$ (right panel), obtained with different smoothing techniques. LOESS: locally weighted regression, GPR: Gaussian process regression.}
    \label{fig:smoothing_comparison}
\end{figure}

Smoothing techniques can be used to determine an estimate of $\Exp_{\mathbf{X}|\mathbf{X}_{\upsilon}=\mathbf{x}_{\upsilon}}\left[u(\mathbf{X},a)\right]$ based on the samples of $u(\mathbf{X},a)$. These techniques include the moving average, linear regression, locally weighted regression (LOESS), Gaussian process regression (GPR) and kernel regression \cite{storlie2008multiple,strong2014estimating,Iooss2015a}.
Figure \ref{fig:smoothing_comparison} illustrates different smoothing estimators $\mathcal{S}(x_i,a)\approx \Exp_{\mathbf{X}|X_i=x_i}\left[u(\mathbf{X},a)\right]$ for the working example for inputs $M$ and $R_1$ with $a=1$. They all give comparable results and only differ at the boundary, where the accuracy is not critical. 
Smoothing becomes more challenging with an increasing number of variables in $\upsilon$, but for the most important and common case of a single input, it is rather straightforward if the number of samples is sufficiently high. 

With the estimator $\mathcal{S}(\mathbf{x}_{\upsilon},a)$, it is straightforward to evaluate the information value by a (quasi-)Monte Carlo (MC) estimator of the expectation in the first line of Eq. \ref{eq:IV}. Ideally, the same MC samples of $\mathbf{X}_{\upsilon}$ are used as in establishing the smoothing estimator, because it can be expected that $\mathcal{S}$ is more accurate at the location of these samples. 

The information value can be estimated as 
\begin{equation}
	\label{eq:IV_MCS}
		V_{\mathbf{X}_{\upsilon}} \approx \frac{1}{n_{MC}}\sum_{k=1}^{n_{MC}} \max_a \mathcal{S}\left(\mathbf{x}_{\upsilon}^{(k)},a\right)  - \mathcal{S}\left(\mathbf{x}_{\upsilon}^{(k)},a_{opt}\right), 
\end{equation}
wherein $\mathbf{x}_{\upsilon}^{(k)}$ is the $k$th MC sample of $\mathbf{X}_{\upsilon}$

An alternative MC estimator is obtained from the second line of Eq. \ref{eq:IV} as
\begin{equation}
	\label{eq:IV_MCS_2}
	\begin{split}
	a_{opt|\mathbf{X}_{\upsilon}}^{(k)} &= \arg \max \mathcal{S}\left(\mathbf{x}_{\upsilon}^{(k)},a\right), \\
	V_{\mathbf{X}_{\upsilon}} &\approx \frac{1}{n_{MC}}\sum_{k=1}^{n_{MC}} u\left(\mathbf{x}^{(k)},a_{opt|\mathbf{X}_{\upsilon}}^{(k)} \right)  - u\left(\mathbf{x}^{(k)},a_{opt}\right).
	\end{split}
\end{equation}

The advantage of this second estimator is that the smoothing estimator is only used in determining the optimal decision, but not in evaluating the expected utility associated with the decision. Hence, this estimator is preferable. 
The results presented in Table \ref{tab:IV_example1} were evaluated with this estimator using $10^5$ Monte Carlo samples and LOESS regression for smoothing.



\subsection{Continuous decision alternatives}
For continuous decision alternatives, the information value can also be determined by Eqs. \ref{eq:IV_MCS} or \ref{eq:IV_MCS_2}. However, the smoothing estimator $\mathcal{S}$ must now also be a continuous function of $a$.
To obtain this smoothing estimator, we propose to artificially treat $a$ as a random variable $A$, and then determine a smoothing estimator of the conditional expectation $\mathcal{S}\left(\mathbf{x}_{\upsilon},a\right) \approx \Exp_{\mathbf{X}|\mathbf{X}_{\upsilon}=\mathbf{x}_{\upsilon},A=a}\left[u(\mathbf{X},a)\right]$. 
While arbitrary distributions can be used for $a$, we recommend using a uniform distribution with bounds $a_{min} \le a \le a_{max}$. These bounds should cover the conditionally optimum $a$ given $\mathbf{X}_{\upsilon}$ for all samples of $\mathbf{X}_{\upsilon}$.

As in the discrete case, Eq. \ref{eq:IV_MCS_2} is expected to give the more accurate estimate. However, for a continuous $a$, the terms
$u\left(\mathbf{x}^{(k)},a_{opt|\mathbf{X}_{\upsilon}}^{(k)} \right)$ and $u\left(\mathbf{x}^{(k)},a_{opt}\right)$ are not already available from the initial Monte Carlo analysis. 
When the utility function is of the form $u(\mathbf{X},a)=u\left(\mathcal{Y}(\mathbf{X}),a\right)$, 
these terms can be computed cheaply from the samples $y^{(k)}=\mathcal{Y}(\mathbf{X}^{(k)})$. In other cases, either the model is reevaluated for all samples with $a_{opt|\mathbf{X}_{\upsilon}}^{(k)}$ or Eq. \ref{eq:IV_MCS} must be used. 

Because a continuous optimization problem needs to be solved in Eqs. \ref{eq:IV_MCS} or \ref{eq:IV_MCS_2}, ideally the smoothing function $\mathcal{S}\left(\mathbf{x}_{\upsilon},a\right)$ is differentiable with respect to $a$. E.g., the moving average is not suitable here. Also, proper regularization is important for regression approaches, to avoid that $\mathcal{S}\left(\mathbf{x}_{\upsilon},a\right)$ has local maxima. 

To limit the number of optimization operations, instead of maximizing $\mathcal{S}\left(\mathbf{x}_{\upsilon}^{(k)},a\right)$ for every sample in Eq. \ref{eq:IV_MCS}, one can solve it for selected values of $\mathbf{X}_{\upsilon}$ and then interpolate the resulting values to obtain a function $S_{opt|\mathbf{X}_{\upsilon}}(\mathbf{x}_{\upsilon}) \approx a_{opt|\mathbf{X}_\upsilon}(\mathbf{x}_{\upsilon})$. Eq. \ref{eq:IV_MCS} then becomes 
\begin{equation}
	\label{eq:IV_MCS_alt}
		V_{\mathbf{X}_{\upsilon}} \approx \frac{1}{n_{MC}}\sum_{k=1}^{n_{MC}} \mathcal{S}\left(\mathbf{x}_{\upsilon}^{(k)},S_{opt|\mathbf{X}_{\upsilon}}(\mathbf{x}_{\upsilon}^{(k)})\right)  - \mathcal{S}\left(\mathbf{x}_{\upsilon}^{(k)},a_{opt}\right).
\end{equation}

In the working example, the information value has been estimated by sampling $a$ uniformly between $4$ and $20$. We utilize Eq. \ref{eq:IV_MCS_2} and determine $a_{opt|\mathbf{X}_{\upsilon}}^{(k)}$ using semi-analytical derivatives of the conditionally expected utility functions. 
Smoothing is performed by means of LOESS.

\textit{Remark}. When the utility function is of the form $u(\mathbf{X},a)=u\left(\mathcal{Y}(\mathbf{X}),a\right)$, it is often possible to derive $a_{opt|\mathbf{X}_\upsilon}$ in function of an expectation over $\mathbf{X}$ that does not involve $a$. Examples are the quadratic utility function of Eq. \ref{eq:u_quadratic} or the LINEX utility function \ref{eq:u_LINEX} introduced in Section \ref{sec:continuous_decisions}. For the quadratic utility function, the conditionally optimal $a$ is simply $a_{opt|\mathbf{X}_{\upsilon}}=\Exp_{\mathbf{X}|\mathbf{X}_{\upsilon}}[\mathcal{Y}(\mathbf{X})]$. For the LINEX function, the optimal decision is
\begin{equation}
	\label{eq:a_opt_LINEX_conditional}
	a_{opt|\mathbf{X}_{\upsilon}} = \frac{1}{\gamma}\ln  \Exp_{\mathbf{X}|\mathbf{X}_{\upsilon}}  \left\{ \exp\left[\gamma \mathcal{Y}(\mathbf{X})\right]\right\}.
\end{equation}

In this case, a smoothing estimator $\mathcal{S}_{LINEX}(\mathbf{x}_{\upsilon}) \approx \Exp_{\mathbf{X}|\mathbf{X}_{\upsilon}=\mathbf{x}_{\upsilon}} \left\{ \exp\left[\gamma \mathcal{Y}(\mathbf{X})\right]\right\}$ can be derived based on the available Monte Carlo samples of $\mathbf{X}$ and $\mathcal{Y}(\mathbf{X})$ and the conditionally optimal decision is then obtained for each sample $k$ as
\begin{equation}
	\label{eq:a_opt_LINEX_conditional_approx}
	a_{opt|\mathbf{X}_{\upsilon}}^{(k)} = \frac{1}{\gamma}\ln  \mathcal{S}_{LINEX}\left(\mathbf{x}_{\upsilon}^{(k)}\right). 
\end{equation}
Inserting into the second line of Eq. \ref{eq:IV_MCS_2} results in the information value. By requiring smoothing only over $\mathbf{X}_{\upsilon}$, and not also over $a$ and avoiding the numerical optimization of $a$, this approach has higher accuracy (or requires lower computational cost to reach the same accuracy). 









.

 \subsection{Rare events} \label{sec:rare_events}
When dealing with models for risk- and reliability assessment, the model output is often a rare event $F$ (typically a failure event). In these cases, a crude Monte Carlo approach is inefficient because a large number of samples are necessary to obtain a few realizations of the rare event. This also holds for quasi-random sequences. Therefore, dedicated rare event estimation methods, such as importance-sampling-based methods or sequential MC techniques, are commonly used to determine $\Pr(F)$. Hence an alternative approach is needed to determine the information value in this case, which we proposed in \cite{straub2022decision}.


For decisions involving rare events, the utility function typically can be written as $u(\mathbf{X},a)=u(I_F(\mathbf{X},a),a)$, wherein $I_F(\mathbf{x},a)$ is the indicator function, which results in $1$ if the specific combination of $\mathbf{x}$ and $a$ corresponds to a failed system and $0$ otherwise. 
For rare failure events, the coefficient of variation of $I_F(X,a)$, and hence of $u(I_F(\mathbf{X},a),a)$, is large. Therefore, a crude Monte Carlo estimate of $\Exp_{\mathbf{X}|X_i}\left[u(I_F(\mathbf{X},a),a)\right]$ is inefficient. However, one can write 
\begin{equation}
	\label{eq:EU_rare}
	\Exp_{\mathbf{X}|\mathbf{X}_{\upsilon}}\left[u(I_F(\mathbf{X},a),a)\right] = u\left(1,a\right)\Pr\left(F|\mathbf{X}_{\upsilon}=\mathbf{x}_{\upsilon}\cap a\right) + u\left(0,a\right)\left(1-\\Pr\left(F|\mathbf{X}_{\upsilon}=\mathbf{x}_{\upsilon}\cap a\right) \right),
\end{equation}
wherein $u\left(1,a\right)$ is the utility associated with decision $a$ and failure $F$ of the system and $u\left(0,a\right)$ is the utility associated with decision $a$ and survival of the system. 

Hence, the conditional expectation of the utility given $\mathbf{X}_{\upsilon}$ and $a$ is a function of the conditional probability of failure. 
\cite{straub2022decision}, based on ideas from \cite{li2019global}, proposes an approach to determine the conditional $\Pr\left(F|\mathbf{X}_{\upsilon}=\mathbf{x}_{\upsilon}\right)$ based on failure samples, i.e., samples of $\mathbf{X}$ that correspond to the event $F$. Such samples are a side-product of any sampling-based reliability method, so that Eq. \ref{eq:EU_rare} can be evaluated without additional model runs. We refer to \cite{straub2022decision} for details. 




\subsection{Computation of the expected values of perfect information \textit{EVPI}}
\label{sec:EVPI_computation}

The computation of the EVPI is straightforward and cheap if the optimal decision for deterministic $\mathbf{X}$, 
\begin{equation} \label{eq:a_opt_EVPI}
	a_{opt|\mathbf{X}}^{(k)} = \arg \max_a u\left(\mathbf{x}^{(k)},a\right),
\end{equation}
can be obtained for every Monte Carlo sample $k$ with limited effort. In particular, for discrete decisions, $a_{opt|\mathbf{X}}^{(k)}$ is readily available. In these cases, the EVPI is estimated as 
\begin{equation}
	\label{eq:EVPI_MCS}
		EVPI \approx \frac{1}{n_{MC}}\sum_{k=1}^{n_{MC}} u\left(\mathbf{x}^{(k)},a_{opt|\mathbf{X}}^{(k)} \right)  - u\left(\mathbf{x}^{(k)},a_{opt}\right).
\end{equation}

For continuous decisions, $a_{opt|\mathbf{X}}^{(k)}$  can typically be computed cheaply or even analytically when the utility function is of the form $u(\mathbf{X},a)=u\left(\mathcal{Y}(\mathbf{X}),a\right)$, i.e., when the QoI does not depend on $a$. 
For the quadratic utility function, Eq. \ref{eq:u_quadratic}, and the LINEX utility function of Eq. \ref{eq:u_LINEX} it is $a_{opt|\mathbf{X}}=\mathcal{Y}(\mathbf{X})$. 

For the more general form $u(\mathbf{X},a)=u\left(\mathcal{Y}(\mathbf{X},a),a\right)$, one might still be able to determine the deterministic optimum $a_{opt|\mathbf{X}}^{(k)}$ cheaply. An example is the continuous decision case of the working example, Section \ref{sec:working_example_continuous}. Here it is 
\begin{equation} \label{eq:opt_decision_working_continuous_deterministic}
a_{opt|\mathbf{X}}^{(k)}=\frac{S^{(k)}}{X_R^{(k)}}.
\end{equation}

If the evaluation of $a_{opt|\mathbf{X}}^{(k)}$ requires additional computationally costly model evaluations, it may not be worthwhile to spend extra efforts to compute the EVPI in practical applications.

\subsection{Computation of the expected values of the perfect model \textit{EVPM}} \label{sec:EVPM_calculation}

Computation of the EVPM is more challenging than of the EVPI. One needs to identify the optimal decision under complete knowledge of the epistemic uncertainty $\mathbf{X}_e$. Due to the aleatory uncertainty, this remains an optimization problem under uncertainty.

If it is possible to eliminate aleatory uncertainty at the onset and work with $\mathcal{Y}_e(\mathbf{X}_e,a)$, one is left with only epistemic uncertainty, and the EVPM is equal to the EVPI and can be evaluated following Section \ref{sec:EVPI_computation}. For discrete problems, this is straightforward. 
For continuous problems, determining the optimal decision in function of $\mathbf{X}_e$ can be more challenging than determining the optimal decision in function of all input variables $\mathbf{X}$. 

In the case of the working example, the solution of Eq. \ref{eq:a_opt_EVPI} in function of $\mathbf{X}_e$ is not as simple as Eq. \ref{eq:opt_decision_working_continuous_deterministic}. Nevertheless, an analytical solution can be found:
\begin{equation}
    a_{opt|\mathbf{X}_e}^{(k)} = \frac{-\ln\left[-\ln\left(1-\frac{10^6}{C_F^{(k)} X_R^{(k)}}\right)\right]+M^{(k)}}{X_R^{(k)}}
\end{equation}

\section{Applications}

We present two real-world applications in which we used information values in the modeling and decision-making process. 

\subsection{Cost-benefit analysis of flood protection}
\label{sec:application_flood}

At the Bavarian Danube River in Germany, potential locations for large-scale flood detention basins were identified \citep{DeVos2021}. To assess their economic efficiency, a benefit-cost analysis is conducted. For that purpose, a probabilistic model $\mathcal{Y}$ was developed that estimates the expected annual flood risk within the study area. This model integrates pre-computed models and simulation outputs that represent climatic, hydrological, hydraulic, geo-hydraulic and economic processes within a hierarchical Monte Carlo framework \citep{Ruf2024, Ruf2023}. The study is ongoing; we show here results from an intermediate model that will still undergo changes before the final assessment.

The model $\mathcal{Y}(\mathbf{X}=[\mathbf{X}_e,\mathbf{X}_a],a)$ predicts the flood damage. It incorporates both aleatory uncertainty, stemming from the inherent randomness of the flood process represented by random variables $\mathbf{X}_a$, as well as (epistemic) model uncertainty represented by $\mathbf{X}_e$. 
Examples of aleatory random variables incorporated in the model include the flood scenarios, which were extracted from a climate simulation, and the performance of the individual dike sections during flood events. In this way, the model includes hundreds of aleatory random variables. 

The model $\mathcal{Y}_e(\mathbf{X}_e,a)$ is the expectation of the flood damages over the aleatory random variables, conditional on the epistemic uncertainty factors $\mathbf{X}_e$. I.e., it provides the risk associated with a flood event in function of $\mathbf{X}_e$ and the decision $a$.
Given the large number of aleatory random variables and the strong non-linearity of the model result with respect to $\mathbf{X}_a$, we evaluate this conditional expectation through a Monte Carlo analysis with $n_{MC_a}=3600$ samples:
\begin{equation}
\begin{split}       \mathcal{Y}_e(\mathbf{x}_e,a)&=\Exp_{\mathbf{X}_a|\mathbf{X}_e=\mathbf{x}_e}[\mathcal{Y}([\mathbf{x}_e,\mathbf{X}_a],a)]\\
    &\approx  \frac{1}{n_{MC_a}}\sum_{j=1}^{n_{MC_a}} \mathcal{Y}([\mathbf{x}_e,\mathbf{x}_a^{(j)}],a)
\end{split}
\label{eq:E_X_e}
\end{equation}

The total risk $R(\mathbf{X}_e,a)$ over a time period $T$ is defined as the sum of the discounted annual risks:
\begin{equation}
R(\mathbf{X}_e,a)= \frac{1-(1+r_d)^{-T}}{r_d} \cdot \lambda_F \cdot \mathcal{Y}_e(\mathbf{x}_e,a)
\end{equation}
$\lambda_F=0.02$/yr is the frequency of relevant flood events and $r_d=2\,\%$ is the annual discount rate. $T$ is taken as the operational lifetime of retention basins, $T=100$yr.

\begin{figure}[h!] \centering
        \includegraphics[width=.75\textwidth]{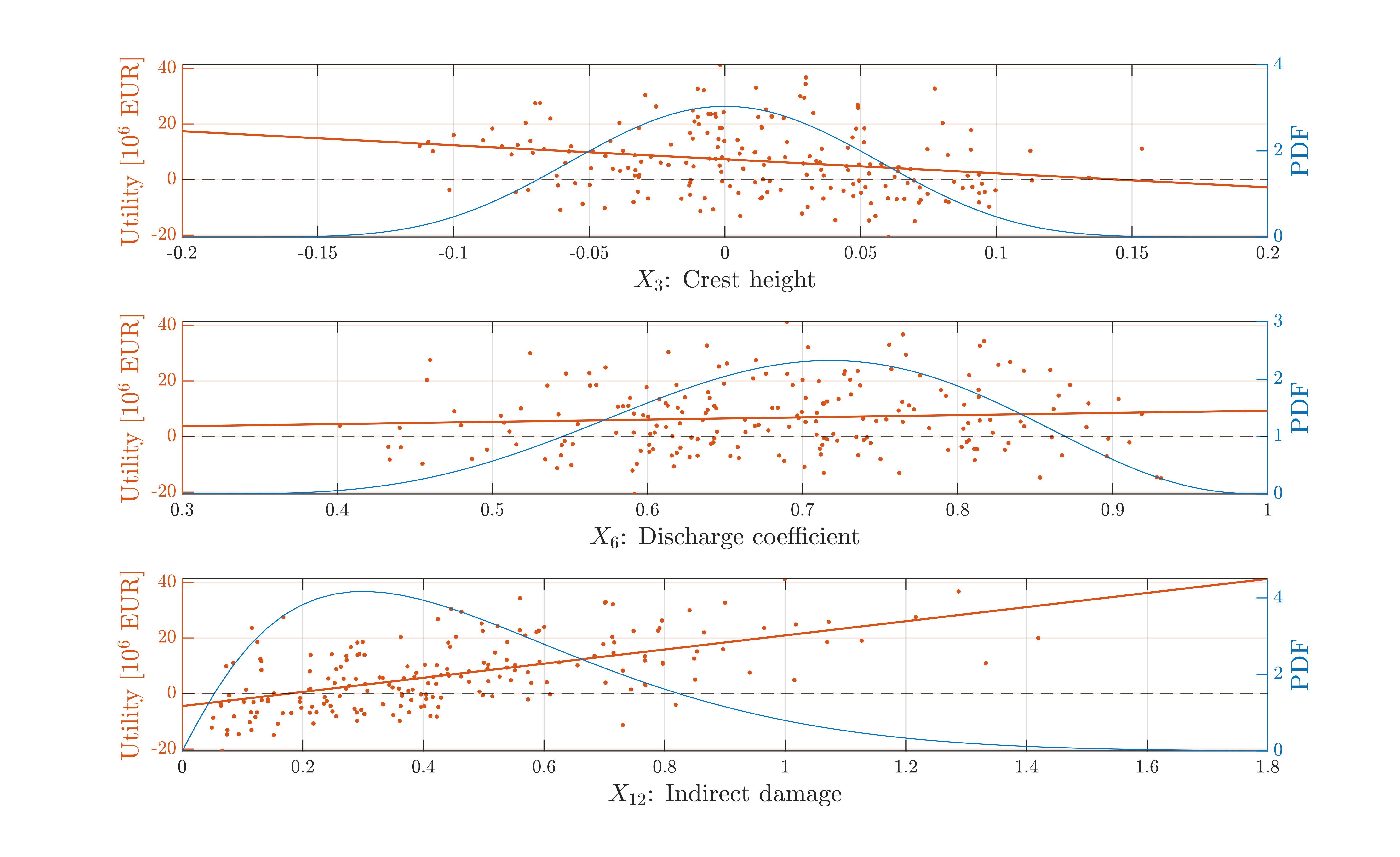}
        \caption{In red: Conditional expectation of the utility function given the three considered model uncertainty factors $X_3$, $X_6$ and $X_{12}$ obtained with a linear regression. In blue: underlying PDF of factors.}
        \label{fig:A1.2}
\end{figure}

The epistemic uncertainty is represented by $14$ random variables $\mathbf{X}_e$, modeled by beta distributions. For illustration purposes, we here focus on three selected factors, namely the uncertainty in the dike crest heights  $X_{3}$, modeled by an additive factor; the discharge coefficient used in the calculation of the outflow during a dike breach $X_{6}$, and the indirect damage $X_{{12}}$, represented as a proportion of direct damage. The distributions of these three model uncertainty factors are illustrated in blue in Figure \ref{fig:A1.2}.

For illustration, we examine here the hypothetical binary decision $a$ on the implementation of a single detention basin. Utility is defined as the net sum of costs and benefits of the measure, where benefits are equal to the risk reduction. The resulting utility function is 
\begin{equation}
    u(\mathbf{X}_e,a)=
    \begin{cases}
    -C_M-R(\mathbf{X}_e,a_M)+R(\mathbf{X}_e,a_0),& \text{if } a=a_M,\\
    0,              & \text{if }a=a_0.
\end{cases}
\end{equation}
where $a_M$ is the decision to implement and $a_0$ is the decision not to implement the detention basin. $C_M$ is the discounted sum of construction and maintenance costs for the detention basin $M$ and is modeled as deterministic. $R(\mathbf{X}_e,a_M)$ and $R(\mathbf{X}_e,a_0)$ are the risks associated with the implementation of the basin and without it, respectively. Following the model, the optimal decision is to implement the detention basin if its expected utility $\Exp_{\mathbf{X}_e}[u(\mathbf{X}_e,a_M))]$ is positive, which corresponds to its benefit-cost ratio exceeding $one$, $\operatorname{BCR}>1$; otherwise the measure should not be implemented ($a=a_0$):
\begin{align}
    a_{opt}=
    \begin{cases}
      a_0, & \text{if } \Exp_{\mathbf{X}_e}[u(\mathbf{X}_e,a_M)]\leq 0\\
      a_M, & \text{if } \Exp_{\mathbf{X}_e}[u(\mathbf{X}_e,a_M)] > 0
    \end{cases}  
\end{align}

For the detention basin under consideration, the expected utility is estimated as 
\begin{equation}
    \Exp_\mathbf{X}[u(\mathbf{X},a_M)]=-C_M+\Exp_\mathbf{X}[R_0(\mathbf{X})-R_M(\mathbf{X})]= - 41.8  \cdot 10^6 \text{\euro} + 48.5 \cdot 10^6 \text{\euro} =6.8 \cdot 10^6 \text{\euro}.
\end{equation}
Consequently, the optimal decision without additional information is to implement the detention basin, $a_M$. To assess the sensitivity of the expected utility to variations in the model parameters, we evaluate the information values $V_{X_i}$. For comparison, we also evaluate the first-order Sobol' indices $S_{X_i}$ assuming the a-priori optimal decision $a_M$. All metrics are estimated from $200$ evaluations of the utility function $u(\mathbf{X}_e,a_M)$ with Monte Carlo samples of $\mathbf{X}_e$. To approximate $\Exp_{\mathbf{X}|X_v=x_v}[u(\mathbf{X},a_M)]$ based on the samples of $u(\mathbf{X}_e,a_M)$, a one-dimensional linear regression $\mathcal{S}$ is applied. The result for the three selected factors is shown as a red line in Figure \ref{fig:A1.2}. 

The resulting sensitivity indices are summarized in Table \ref{tab_evppi_sob}.  
The decision is sensitive to the uncertainty in factor $X_{12}$. 
By contrast, the uncertainty in parameters $X_3$ and $X_6$ has minimal influence on the optimal decision. The first-order Sobol' indices show a similar picture, although a less pronounced one. In particular, they do not clearly show that the uncertainty in $X_3$ has a negligible influence on the decision. Furthermore, the information value has the benefit of being interpretable in an absolute sense, i.e., the decision makers can decide if it is worthwile to spend additional time and money on reducing the uncertainty in $X_{12}$ if the potential benefit is capped at $215 \cdot 10^3$\euro.

\begin{table}[h]
	\centering
	\caption{Information value and first order Sobol' indices for selected epistemic uncertainty factors $X_3$, $X_6$ and $X_{12}$.}
	\label{tab_evppi_sob}
	\small{
	\begin{tabular}{ccc}
	\hline
	Uncertainty & Information Value [€] & First order Sobol' index \\ \hline
        $X_3$  & $0.2 \cdot 10^3$ & $0.05$ \\
        $X_6$  &  $0$ & $0.01$\\
        $X_{12}$  &  $215 \cdot 10^3$ &  $0.34$ \\
        \hline
	\end{tabular}
}
\end{table}

\begin{figure}[h] \centering
   \includegraphics[width=0.8\linewidth]{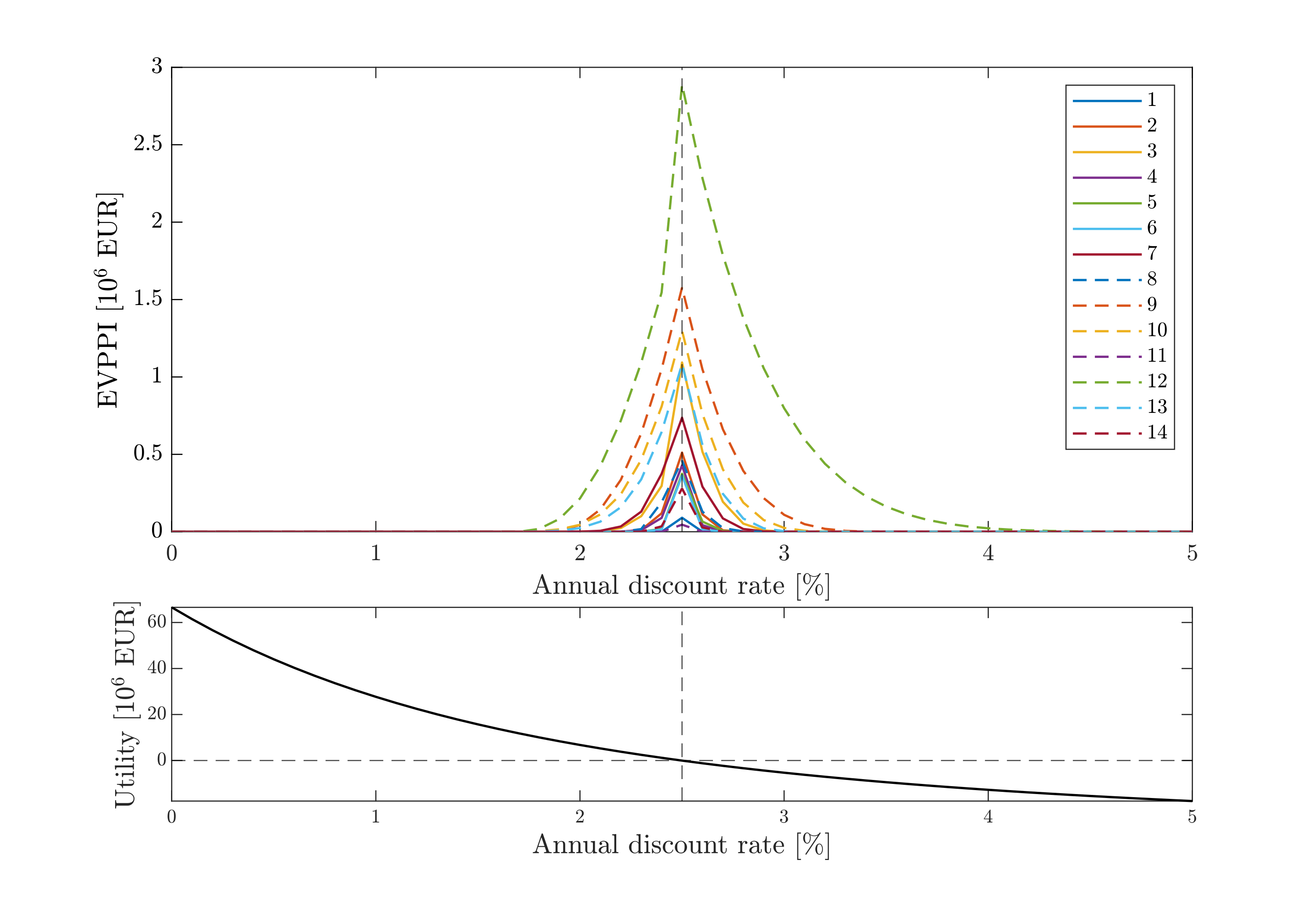}
   \caption{Information Value EVPPI of all 14 model uncertainty factors (upper panel) and utility (lower panel) in function of underlying annual discount rate.}
   \label{fig:A1.1}
\end{figure}

The results presented above are based on an annual discount rate of $d=2\%$. In the following, we investigate the effect of changing $d$ to further illustrate the decision-oriented nature of the information value $V_{X_i}$. While the benefits of a flood risk mitigation measure are distributed over the operational lifetime of the measure, 
the majority of costs occur during the initial construction period. Consequently, a higher discount rate $d$ reduces the utility of the mitigation measure but does not alter the cost $C_M$ much. If a discount rate of $d=1\%$ is used, the expected utility increases to $\Exp_\mathbf{X}[u(\mathbf{X},a_M)]=27.7 \cdot 10^6 \text{\euro}$ and all $V_{X_i}$ values decrease to $zero$\footnote{The $zero$ values are computed with the MC approximationd. The exact values of the $V_{X_i}$ are larger than zero, but negligible.}. This implies that learning the value of $X_{12}$ would not alter the optimal decision, even if knowledge about $X_{12}$ can significantly influence the estimate of the utility. This is reflected by the first-order Sobol’ indices, which remain constant across varying values of $d$ even if the decision sensitities change drastically.
The effect of varying the discount rate $d$ on the information values $V_{X_i}$ is illustrated in the upper panel of Figure \ref{fig:A1.1}. As the expected utility moves further away from the decision threshold $u(\mathbf{X}, a_M) = 0$, the uncertainty in the parameters becomes less relevant to the decision-making process. The absolute values of the $V_i$ values are largest at the point when the expected utility associated with the decision $a_M$ is $zero$. This happens at around $d=2.5$. In this case, the decision is most likely to change when learning about any of the model uncertainties. The relative contribution of the different factors remains similar across varying discount factor values.

\subsection{Selection of a site for nuclear waste disposal}
\label{sec:application_nuclearWaste}

In the search for a suitable location to construct a deep geological repository for radioactive waste in Switzerland, three potential siting regions ($a\in\{1,2,3\}$) were identified 
during stage 2 of the site selection process \cite{ensi2018}
by the National Cooperative for the Disposal of Radioactive Waste (NAGRA). The site selection process is involved and comprises a total of 13 criteria \cite{nagra2022}. This application example focuses solely on the criterion \textit{Erosion} \cite{nagra2024}, which is connected to the long-term safety and stability of the natural barrier properties. For this criterion, the remaining overburden above the repository, as indicator for total erosion, is modeled for different time points in the future, taking geological, fluvial and glacial processes into account. 
A probabilistic model was established based on an extensive and iterative process, which combines models for fluvial incision, evolution of local topography and glacial overdeepening.
The model for fluvial incision returns the local erosion base of the main river, upon which the evaluation of local topography and glacial overdeepening is based.
Model parameters $\mathbf{X}$, including their uncertainties, were estimated using a systematic expert elicitation process \cite{dias_shelf_2018}.
The models enable the computation of the probability of exposure for all siting regions $p_F(\mathbf{X},a)$ at future times, where exposure is associated with negative or zero remaining overburden. 
For the purpose of sensitivity analysis, we consider that the siting region with the smallest probability of exposure $p_F$ (due to erosion) after one million years is preferred with respect to the Erosion criterion. 
Note that this approach does not reflect the actual site selection procedure, but supports the erosion assessment.

To derive the utility function associated with the Erosion criteria, the simplifying assumption was made that there are no consequences if the repository is not exposed within one million years and that there are fixed (but unknown) consequences if the remaining overburden is reduced to zero. 
Differences in costs between the different siting regions are not considered.
It follows that the utility associated with siting region $a$ and probabilistic parameters $\mathbf{X}$ is proportional to the negative probability of exposure:
\begin{equation}
	\label{eq:utility_nuclear}
		u(\mathbf{X},a) = - p_F(\mathbf{X},a)
\end{equation}
Note that the probability of exposure $p_F(\mathbf{X},a)$ is the expected value of an indicator function (c.f. Section \ref{sec:rare_events}) with respect to aleatory uncertainties. With the applied models, it is possible to compute this probability semi-analytically for given values of $\mathbf{X}$.

The probabilistic model to evaluate the probability of exposure $p_F(\mathbf{X},a)$ at the three siting regions $a$ depends on a large number of uncertain input parameters $\mathbf{X}$. The parameter vector $\mathbf{X}$ includes parameters controlling uplift, erodibility of lithostratigraphic units,  future drainage scenarios, evolution of local topography, future glaciation events and glacial overdeepenings.
For illustration, we again focus on three selected parameters $X_1$, $X_4$ and $X_9$:
The parameter $X_1$ represents the average annual rock uplift at a spatial reference point.
The parameter $X_4$ is the base-level at another spatial reference point.
$X_9$ is a scaling parameter linked to siting region $a=2$ that controls how deep a glacier can incise into rock layers belonging to erodibility class EC4 (according to an internally applied rock classification scheme).

\begin{figure}
    \centering
    \begin{subfigure}{.5\textwidth}
        \centering
        \includegraphics[width=.95\textwidth]{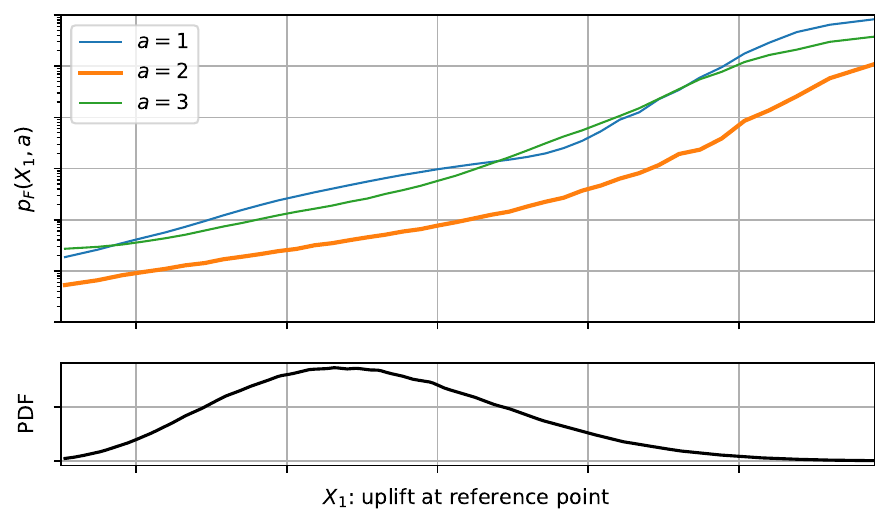}
    \end{subfigure}%
    \begin{subfigure}{.5\textwidth}
        \centering
        \includegraphics[width=.95\textwidth]{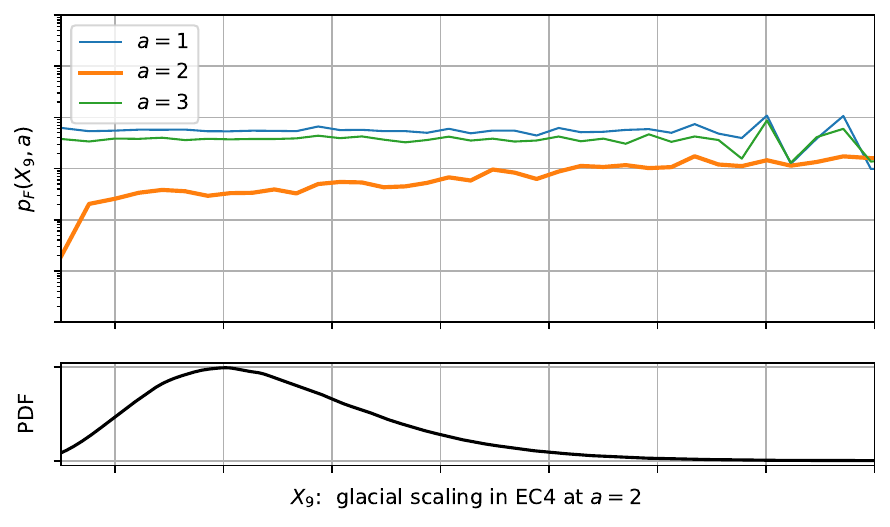}
    \end{subfigure}
    \caption{%
        The probability of exposure $p_F$ at the three siting regions ($a\in\{1,2,3\}$) is shown conditional on $X_1$ (left side) and $X_9$ (right side) in the plots of the upper row.
        In the plots of the lower row, the associated probability density function (PDF) of $X_1$ (left side) and $X_9$ (right side) is shown.
    }
    \label{fig:A2_1}
\end{figure}

The full Monte Carlo analysis finds that the optimal siting region is "N{\"o}rdlich L{\"a}gern", $a_{opt}=2$, since it has the lowest probability of exposure. This site is also the final choice made by NAGRA considering all decision criteria \cite{muller_status_2024}. 
The probability of exposure $p_F$ at the three siting regions is shown in Figure~\ref{fig:A2_1} conditional on $X_1$ (left side) and $X_9$ (right side).
For $X_1$, the conditional probability of exposure clearly increases with increasing values of $X_1$ at all three siting regions. 
Thus, the first-order Sobol' indices for $X_1$ are large for each siting region; i.e., at each siting region parameter $X_1$ clearly contributes to the output variance of the model. 
Nevertheless, the information value for $X_1$ equals \emph{zero} ($V_{X_1}=0$), as the probability of exposure is always smallest at siting region $a=2$.
Consequently, learning the value of $X_1$ will not change the optimal decision even though knowledge about $X_1$ can significantly impact the estimate of the probability of exposure.
For $X_9$, the probability of exposure is by definition independent of $X_9$ at the siting regions $a=1$ and $a=3$. The probability of exposure at siting region $a=2$ increases slightly with increasing values of $X_9$.
Also for parameter $X_9$, the probability of exposure is always smallest at siting region $a=2$ (except for some statistical noise in the upper tail of $X_9$) and, therefore, $V_{X_9}=0$.

\begin{figure}
    \centering
    \begin{subfigure}{.5\textwidth}
        \centering
        \includegraphics[width=.95\textwidth]{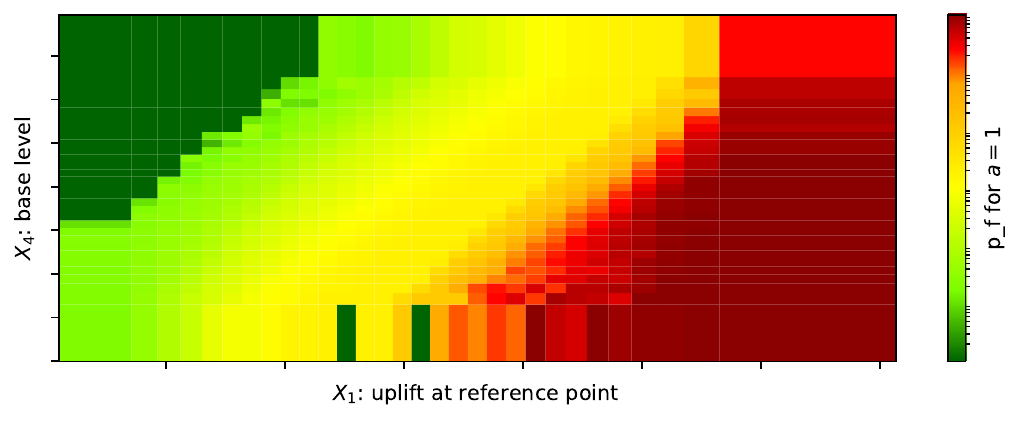}
        \caption{$p_F$ for $a=1$}
    \end{subfigure}%
    \begin{subfigure}{.5\textwidth}
        \centering
        \includegraphics[width=.95\textwidth]{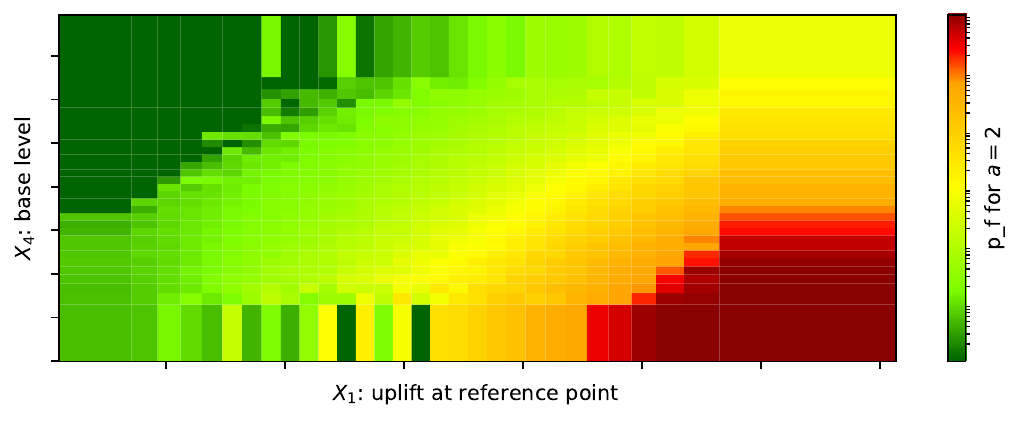}
        \caption{$p_F$ for $a=2$}
    \end{subfigure}
    
    \vspace{1em}
    \begin{subfigure}{.5\textwidth}
        \centering
        \includegraphics[width=.95\textwidth]{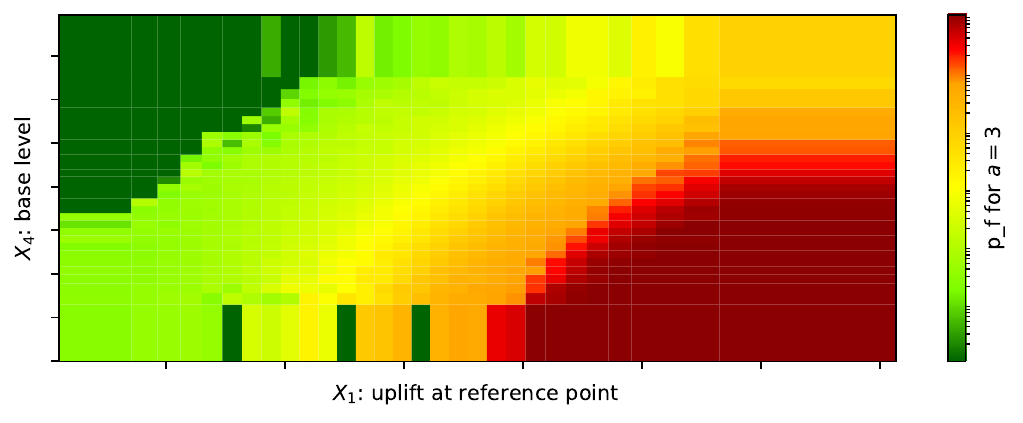}
        \caption{$p_F$ for $a=3$}
    \end{subfigure}%
    \begin{subfigure}{.5\textwidth}
        \centering
        \includegraphics[width=.95\textwidth]{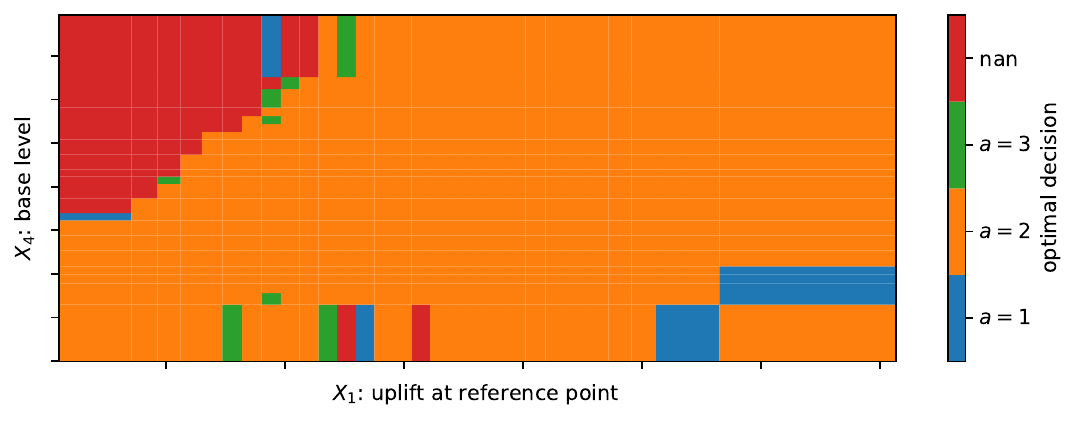}
        \caption{optimal decision conditional $X_1$ and $X_4$}
    \end{subfigure}
    \caption{%
        (a), (b) and (c) show the probability of exposure conditional on $X_1$ and $X_4$ for the siting regions $a=1$, $a=2$ and $a=3$.
        (c) shows the siting region that is optimal conditional on knowledge of $X_1$ and $X_4$. 
        For fields in (c) that are colored in red, an optimal siting region could not be identified, due to small probabilities of getting values of $X_1$ and $X_4$ that fall in these fields ($X_1$ and $X_4$ exhibit a positive dependence structure).
    }
    \label{fig:A2_2}
\end{figure}

Figure~\ref{fig:A2_2} illustrates how the probability of exposure depends jointly on both $X_1$ and $X_4$. 
Large values of $X_1$ in combination with small values of $X_4$ result in large values for the probability of exposure at all three siting regions. 
Consequently, acquiring information about both $X_1$ and $X_4$ can considerably reduce the uncertainty about the probability of exposure.
Nevertheless, the information value $V_{\{X_1,X_4\}}$ of the parameter group $\{X_1,X_4\}$ is \emph{zero} with respect to identifying the optimal siting region: For all combinations of values of $X_1$ and $X_4$ (not accounting for statistical noise), the siting region $a=2$ remains the optimal siting region.

In this application, the information value of all uncertain factors $\mathbf{X}$ is zero. This demonstrates to the decision maker and stakeholders that the model is sufficiently informative and that collecting additional information on the parameters will not change the decision.

\section{Concluding remarks}
Information value (IV) is a powerful sensitivity measure in the context of engineering and environmental decision-making. 
In this article, we showcase the usefulness and simplicity of this sensitivity measure and present the modeling and computational steps needed for its evaluation. 
Along the way, we also include a set of novel contributions, summarized in the following.

We investigate the separation between aleatory and epistemic uncertainty into the sensitivity analysis and discuss its importance for interpretation and computations. Related to this, we introduce the expected value of the perfect model (EVPM) and propose to use this for normalizing IV, which is a natural way. 

We present the computation of the IV for applications in which the decision is characterized by a continuous parameter, which is often the case in engineering design situations. We discuss that this can lead to additional computational challenges, such as the choice of a suitable smoothing function that facilitates optimization of $a$, which we did not address in detail and leave for further work.

We present two real-world applications of the sensitivity measures that demonstrate their benefits for engineering and environmental decision-making. These applications highlight a particular benefit of the IV over other sensitivity measures, namely that it can be interpreted in an absolute sense. By providing an upper bound to the potential value of collecting information on an input quantity, it facilitates the adaptive modeling process in which a model can be refined until the possible benefit of removing the remaining uncertainty is low compared to the cost of doing so.

Furthermore, we also propose to use the probability of a decision change as an additional sensitivity measure that we found to be useful in practical applications for communicating the effect of a particular input uncertainty. If this probability is small, the decision-makers are more likely to accept the uncertainty associated with this input parameter. Its interpretation is even more straightforward than the IV's interpretation, even if the IV is the more relevant measure from a decision-theoretic viewpoint.

In this paper, we present the computation of IV and the EVPM based on the premise that samples of the model inputs and outputs are available and that the IV should be computed based on a mere postprocessing of these samples. In the case of discrete decisions, this is straightforward if the model is evaluated for all decision cases. The only challenge that can arise is if the sensitivity measure should be determined for larger sets of input variables, in which cases the conditional expectation (smoothing) needs to be performed in higher dimensions. The analysis of inputs with sets in the order of 10 input variables using GPR is reported in the literature \cite{strong2014estimating}, but to our knowledge, there currently exists no library that can perform the IV analysis automatically in these cases, and there is a potential in developing such a library.

In the case of continuous decisions, there is a need to develop smoothing methods that allow the robust identification of the optimum decision, as pointed out in the paper. Future work should look into this. Furthermore, for cases with continuous decisions it seems relevant to perform a systematic investigation into the relation between the IV, the Sobol' index and the generalization of the Sobol' index based on the LINEX function proposed in this paper.




\bibliography{library}

\end{document}